\documentstyle[prl,aps,multicol,epsf,rotate,citesort,subeqn]{revtex}

\newcommand{\gl}[1]{Eq. (\ref{#1})}
\newcommand{\gls}[2]{Eqs. (\ref{#1},\ref{#2})}

\def\gtrless{\raise2.5pt\hbox{$>$}\llap{\lower2.5pt\hbox{$<$}}}
\def\gtrapprox{\raise2.5pt\hbox{$>$}\llap{\lower2.5pt\hbox{$\approx$}}}
\newcommand{\bsq}[1]{\begin{subequations}\label{#1}}
\newcommand{\esq}{\end{subequations}}
\newcommand{\beq}[1]{\begin{equation}\label{#1}}
\newcommand{\eeq}{\end{equation}}
\newcommand{\beqa}[1]{\begin{eqnarray}\label{#1}\nonumber}
\newcommand{\eeqa}{\end{eqnarray}} 
\newcommand{\fur}{\qquad\mbox{for }\, } 
\newcommand{\wer}{\qquad\mbox{where }\quad}
\newcommand{\lam}{\lambda}
\newcommand{\vek}[1]{{\bf #1}} 
\renewcommand{\rho}{\varrho}

\begin{document}
\title{
Macromolecular theory of solvation and structure in mixtures of
colloids and polymers} 
\author{M.~Fuchs$^{1}$\footnote{Permanent address: Physik-Department, 
Technische Universit{\"a}t M{\"u}nchen,
85747 Garching, Germany}
 and K.~S.~Schweizer$^{2}$}
\address{$^{1}$ Department of Physics and Astronomy, 
The University of Edinburgh, JCMB King's Buildings,
Mayfield Road, Edinburgh EH9 3JZ, United Kingdom\\
$^{2}$ Departments of Materials Science \& Engineering and Chemistry
and Materials Research Laboratory,
University of Illinois, Urbana, Illinois 61801,USA}
\date{\today}
\maketitle

\begin{abstract} 
The structural and thermodynamic properties of mixtures of colloidal
spheres and non-adsorbing polymer chains are studied within a novel general
two-component macromolecular liquid state approach applicable for all
size asymmetry ratios. 
The dilute limits, when one of the components is at infinite dilution but 
the other concentrated, are presented and compared to field theory and 
models which replace polymer coils with spheres.  
Whereas the derived analytical results compare well, qualitatively and 
quantitatively,
  with mean-field scaling laws where available, important differences
from ``effective sphere'' approaches are found for large polymer sizes or
semi-dilute concentrations.
\end{abstract}

\begin{multicols}{2}

\section{Introduction}

 Mixtures of dispersed spherical particles and non-adsorbing polymers
 may be viewed as a model system for a wide variety of materials
 encountered in food products, biological systems or technological
 applications. In these systems, the ``depletion attraction'' is always
  present because it
 has a purely entropic and universal origin. Its consequences can most
 clearly be studied in mixtures of colloidal hard spheres and polymer
 chains made up of hard units where only the  entropic
 consideration of the 
 packing of particles restricted by steric or excluded volume
 constraints enters.

Because of the fundamental nature of the
depletion attraction it has been studied theoretically since the
pioneering work of Asakura and Oosawa \cite{Asakura54} and
Vrij \cite{Vrij76}. Moreover, its effect on the phase behaviour had
been observed much earlier \cite{Traube25}. The phase diagram of
colloid-polymer mixtures has been constructed using the Asakura-Oosawa
pair potential in an effective one-component
 thermodynamic perturbation calculation \cite{Gast83},
within a two-component dilute polymer
 free volume approach \cite{Lekkerkerker92}, and also with
computer simulations \cite{frenkel} and
\cite{Dijkstra99}, the latter  based on a
specific model originating in
Refs. \cite{Asakura54,Vrij76}. The Asakura-Oosawa model
consists of replacing the polymer coils with effective spheres which
can freely interpenetrate each other but not the colloidal spheres. This
model has been further treated by liquid state theory
\cite{Louis99}, has been extended to non-homogeneous situations
\cite{Schmidt00,Brader00}, and to perturbatively include polymer non-ideality
 \cite{Warren95}. The forces it predicts for dilute
rather large colloidal spheres have been measured directly
\cite{Rudhardt98,verma}, and  phase diagrams for
colloidal spheres appreciably larger than the polymers have been
obtained, and agree semi-quantitatively with theory
\cite{Calderon93,Ilett95,Tuinier99}.

More detailed experiments on the colloidal correlations
\cite{frenkel,tong,Moussaid99,Tuinier00b},
measurements of the second virial coefficient
\cite{Kulkarni99}, and quantitative tests of the phase
diagrams for larger polymer sizes
\cite{Ilett95,Moussaid99,Bodnar97,Dzubiella01,Verhaegh96,Tuinier00b},
have, however, detected polymer correlations which are not contained in
the mentioned approaches. 
Also, for small spherical surfactant micelles \cite{Robb95} the dependence
of phase separation on polymer size is opposite to predictions of
colloid approaches \cite{Gast83,Lekkerkerker92}. There are at least two
reasons for these discrepancies.
 First, polymer coils can deform close to
particles and can thus fit into void spaces more effectively than
spheres. Second, for higher polymer concentrations the coils start to
overlap and the relevant polymer correlation length crosses over from
the coil radius to the size of a mesh in the formed transient network. Both
effects are important if the polymer coils are not negligibly small
compared to the particles, and both are neglected in the described
theoretical approaches. The effects have long been understood from
field-theoretic approaches to polymers in the limit of
dilute colloidal particles. The deformability of
the polymer coils affects the depletion layer of polymer segments
close to particles, the resulting insertion free energies, and the
induced colloid pair interactions
\cite{eisenriegler,eisenriegler2,eisenriegler3,Eisenriegler00,Eisenriegler00b}. For
semi-dilute polymer suspensions the depletion layer and the induced
interactions were obtained for both large and small colloids
\cite{eisenriegler3,Eisenriegler00,Eisenriegler00b,joanny,deGennes79,Odijk,Sear98}. Yet,
except for in a highly idealized
 mean field thermodynamic perturbation calculation to
hard spheres by Schaink and Smit \cite{Schaink97}, polymer
field-theoretical approaches have not been extended to finite colloid
concentrations.

Recently we proposed a macromolecular liquid state theory for mixtures
of arbitrary polymer to colloid size ratios \cite{Fuchs00}, which,
although it is not rigorous for dilute systems, presents a viable and
first principles approach for finite densities. It is unique in its
applicability to all parameter ranges concerning densities and
sizes, and is a macromolecular generalization
\cite{kcur1,kcur2,kcur3} of the interaction site description introduced by
Chandler and Andersen for small rigid molecules \cite{chandler}. This
generalization has proven rather successful for pure, especially dense
polymer systems and polymer alloys.
 Some results for dilute particle mixture systems have been obtained
within a simplification of the approach \cite{avik,avik3}, and could
rationalize several surprising
aspects of measured second virial coefficients of small proteins
\cite{Kulkarni99}. Also, light scattering measurements of
the colloid liquid structure could be described semi-quantitatively
over all length scales without adjustable parameters \cite{Fuchs00}.

In the present manuscript we analyze in detail the low density limits of this
macromolecular approach. The reasons are threefold. First, by looking
at polymer solutions containing few colloidal particles it is possible
to compare with exact field-theoretic results and thus to test
 the approach. Second, by considering dilute polymers in a
hard sphere solvent it is possible to make contact with the previous
Asakura-Oosawa type approaches. Third, in these limits fully
analytical solutions of the non-linear integral equations description
are possible and provide insight into the theory, which also applies
to the higher concentration states.

A conceptually new closure (approximation) for the direct correlation 
function describing
the packing of polymers close to repulsive walls or around hard colloidal
spheres has been introduced in \cite{Fuchs00} which entails a 
medium-ranged
colloid-polymer segment effective interaction. 
To capture the key physics the latter
 is required within the polymer reference interaction
 site model (PRISM)
approach since a preaveraging approximation for the single
polymer chain form factor is employed for tractability reasons.
 In inhomogeneous systems, however, the 
single polymer form factor depends on the distance of the polymer chain
 from interfaces or inhomogeneities.
Considering a fluid of random walk polymers, Gaussian intramolecular
correlations apply. The number of intersections of a
random walk with a plane  scales as $\sqrt{N}$, where
$N$ is the number of steps or polymer repeat units. 
 Without rearrangements, the number of 
contacts with a repulsive wall would scale identically, as the excluded
volume constraint could be satisfied by just mirror inverting
the overlapping polymer strands.  This result follows from PRISM 
with the most simple excluded volume closure (of Percus-Yevick form)
\cite{avik}. Close to the repulsive wall,
however, the (Gaussian) intramolecular polymer correlations differ from 
the ones in the bulk solution as translational entropy can be gained by 
reducing the number of contacts with the wall to ${\cal O}(1)$, as
required for recovery of the ideal gas equation of state from the wall
virial theorem.  
In order to describe the inhomogeneous system with one homogeneous 
polymer intramolecular structure factor, the rearrangements close to a 
colloidal particle need to be captured by an effective colloid-polymer
interaction which extends across the range where the polymer segments 
rearrange. We proposed a molecular  closure convoluting the local (bare)
segmental steric repulsion with a Yukawa weight where the range, or
non-locality length, 
called $\lambda$, is determined from thermodynamic consistency
considerations. 
On the segmental level the assumption of short ranged
effective steric interactions entered into a Percus-Yevick style
approximation \cite{avik,avik3}.

Thermodynamic consistency correlates the structure on local length
scales with long wavelength fluctuations and is a familiar concept in
liquid state theories \cite{hansen}. In the present case, the polymer
chemical potential at infinite dilution, closely connected to the
insertion free energy for adding dilute polymers to a particle fluid,
is used to implement consistency 
as it provides one of the simplest measures of the tendency of
colloids and polymers to mix. Moreover, this quantity also determines
the free volume which is one of the input quantities in the colloid
theory most widely used for large colloid to polymer size ratios
\cite{Lekkerkerker92}

The outline of this paper is as follows:
 In section II the  model of  colloid
polymer-mixtures is presented. Section III describes the solution of the
integral equations in the two low density limits of interest.
The thermodynamic consistency
equations are solved and discussed in section IV. Section V then presents the
results and discussions for the structure and thermodynamics of dilute
colloidal particles in a polymer solvent, while section VI describes the 
opposite case of dilute polymer chains immersed in a hard sphere fluid.
Conclusions are in presented section VII, and three
 appendices contain technical material and a discussion of alternative
 closure approximations.

\section{Model}

The binary mixture shall be described by its (matrix of partial) structure
factors $\hat{S}_{ij}(q)$, where the index $i=1$ indicates
 the polymer and $2$ the colloid  component. A small molecule solvent is
 treated as a background continuum and enters only implicitly via the
 interaction potentials for the polymers and colloids.
In principle, all partial
structure factors are experimentally measurable by (labelling and)
scattering techniques.  The total density fluctuations are decomposed into 
single molecule contributions,
described by a (diagonal) intramolecular form factor, 
$\hat{\omega}_{ij}(q) =\hat{\omega}_{i}(q) \delta_{ij}$, and intermolecular
correlations, $\hat{h}_{ij}(q)$, resulting in:
\beq{m1}
\hat{S}(q) =  \varrho\;  \hat{\omega}(q) + \varrho\;  \hat{h}(q)\; \varrho\; .
\eeq
An obvious matrix notation is used.
The diagonal matrix of densities, $\varrho_{ij} = \varrho_i \delta_{ij}$, gives
the number density of colloidal particles and polymer segments. The
pair decomposable excluded
volume or steric  interaction prevents the particles/ segments from 
overlapping:
\beq{m2}
g_{ij}(r< \frac 12 (\sigma_i + \sigma_j ))=0 \; ,
\eeq
where $\sigma_2=\sigma_c$ is the colloidal hard core diameter and
$\sigma_1=\sigma_p$ is the excluded volume diameter of a single 
polymer repeat unit (segment). The intermolecular pair correlation functions,
$g_{ij}(r)$,  are trivially connected to the total intermolecular, $h_{ij}$,
 correlation functions, $g_{ij}(r) = h_{ij}(r)+1$. Carets in \gl{m1} denote 
Fourier-transformed quantities. 
The total density fluctuations of the interacting fluid are decomposed into the
single molecule fluctuations and in an interaction part 
via a generalized Ornstein-Zernicke, or Chandler-Andersen,  equation
\cite{kcur3,chandler,chandl}: 
\beq{m3}
\hat{S}^{-1}(q)  =  \hat{\omega}^{-1} \varrho^{-1} - \hat{c}(q)\; .
\eeq
In a pre-averaging approximation the single polymer-molecule
 density fluctuations
\beq{m4}
\hat{\omega}_{p} \equiv \omega(q) = \frac 1N \sum_{\alpha\beta}^N 
\; \langle e^{i \vek{q} ( \vek{r}_\alpha - \vek{r}_\beta ) } \rangle \; ,
\eeq
are taken to be known a priori. As the colloidal particle
is assumed to be rigid and solid, it
acts as a point scatterer, $\hat{\omega}_{c}=1$, subject
to the steric restriction, \gl{m2}.  If the effective interaction potentials,
the direct correlation functions, $c_{ij}(r)$ in \gl{m3}, were taken to
be the bare pair potentials, $c_{ij}(r) = - V_{ij}(r)/(k_BT)$, then 
 \gls{m1}{m3} would correspond to the RPA approximation
which is one of the simplest liquid state approximations for (dense)
polymeric and simple fluids. Typically one finds that RPA solutions violate
the excluded volume condition, \gl{m2}.  Integral equations approaches
like  PRISM go beyond the RPA as they enforce the no-overlap condition 
\gl{m2} rigorously and determine the direct correlation functions from
self-consistency equations implementing the (physically) motivated 
expectation that the $c_{ij}(r)$ are short ranged and vanish beyond a few
particle diameters. For the colloidal hard sphere component this corresponds 
to the well established Percus-Yevick (PY) approximation \cite{hansen}:
\beq{m5}
c_{cc}(r>\sigma_c)=0 \; .
\eeq
This closure and the excluded volume constraint, \gl{m2}, 
together with the site-site Ornstein-Zernicke equation, \gl{m3},  
result in a coupling of density fluctuations at different wave vectors,
thus leading to non-linear integral equations with 
a much richer mathematical structure  than the simple RPA  approximation. 
For the model of hard spheres, the only thermodynamic parameter is the
packing fraction,  $\phi_c=\frac{\pi}{6} \varrho_c \sigma_c^3$.

Detailed studies of the PRISM equations for homopolymer solutions and melts
\cite{kcur2,kcur3}
have established that the polymer site-site direct correlation function
to a good approximation decays to zero beyond the polymer repeat unit
size so that a correspondingly simple closure can be enforced: 
\beq{m6}
c_{pp}(r>\sigma_p)=0 \; .
\eeq
Thus, the interaction between polymer macromolecules is made up of pairwise
site-site segmental interactions, which are given by a spherically symmetric
effective potential, which follows from the excluded volume constraint, 
\gl{m2}. Attractive interactions beyond the ''athermal`` model studied here
can be included \cite{kcur3,avik3}.
Site averaged quantities are considered, and therefore specific
 chain-end effects are neglected.

As discussed in the introduction,
 the effective colloid-polymer interaction  extends 
beyond the range of immediate overlap. This arises because of, and
 allows to
accommodate, the change of the polymer conformations close to  colloidal
particles.  As the exact direct correlation function is not known we 
suggested \cite{Fuchs00}
a simple one-parameter extension of the PY closure (m-PY):
\bsq{m7}
\beq{m7a}
\hat{c}_{cp}(q)= \frac{\hat{c}^s_{cp}(q)}{1+q^2\lambda^2}\;,
\quad\mbox{with}\quad  c^s_{cp}(r>\frac{\sigma_c+\sigma_p}{2})=0\;,
\eeq
which enforces excluded volume on the local scale by fixing
$c^s_{cp}(r)$ from the excluded overlap condition, \gl{m2}, and from the
requirement of short-ranged  segmental interactions. 
On physical grounds, because $-k_BT\,c^s_{cp}(r)$ describes interactions 
on the segmental scale, one expects $c^{s}_{cp}(r)$ to be negative (repulsive)
and to exhibit rapid variations (on the segmental length scale)
and smoother ones connected with the colloid size.
In real space 
the closure clearly implies a smearing of the segment-colloid
interactions over the distance $\lambda$:
\beq{m7b}
c_{cp}(\vek{r}) = \int\!\!d^3s\; \frac{1}{4\pi\lambda^2} \; \frac{1}{
|\vek{r}-\vek{s}|}\; 
e^{-|\vek{r}-\vek{s}|/\lambda}\; c^s_{cp}(\vek{s})\;, 
\eeq\esq
where the PY closure for $c^s_{cp}(r)$ can be viewed as describing
 unconnected polymer segments,
 and the (nonlocal) conformational constraints on 
the segment packing (``chain connectivity'') close to colloidal particles
are captured by the  spatial convoulution.

The m-PY closure contains an undetermined parameter, the length $\lambda$,
which can be expected to vary non-trivially with the physical 
system parameters, like densities or size ratio. As it captures the
rearrangements of the polymers strands close to a colloidal particle, its
magnitude should be of the order or smaller than the polymer correlation
length (i.e. radius of gyration, $R_g$ for dilute systems,
 blob diameter or mesh size for
semi-dilute conditions) and/or the colloid size. Also, the polymer 
conformational changes, and hence $\lambda$,
 will depend on the volume taken up by the colloidal 
spheres. In order to achieve  a parameter free a priori description,
thermodynamic consistency shall be enforced te determine 
$\lambda$ uniquely. The implementation of this well known concept within
liquid state theories starts from the observation that the PY
closure \cite{kcur3}, $\lambda=0$, leads to results for
the solution free energies obtained from the compressibility theorem,
$\frac{\partial^2}{\partial \rho_i \partial \rho_j} F^{\rm ex.} =
- k_BT \hat{c}_{ij}(0)$, where $F^{\rm ex.}$ is the excess free energy
per unit volume,
which compare favourably with field theoretic results where available
\cite{avik}.   
Thus, the excess chemical potential for inserting polymers into a hard sphere 
fluid, where $\delta\mu_i = \frac{\partial}{\partial \rho_i} F^{\rm ex.}$, 
as obtained via the ``compressibility'' route provides a 
rather $\lambda$-insensitive reference quantity since it emphasizes
long wavelenth correlations. Especially the limit
for vanishing polymer concentration shall be discussed,
\beq{m8}
N \beta \delta\mu^{\rm (c)}_p|_{\varrho_p=0} = 
- \int_0^{\varrho_c} d\varrho'_c\; N\;
\hat{c}_{cp}(q=0,\varrho_c')|_{\varrho_p=0}\; ,
\eeq
where $\beta=1/(k_BT)$, and the expression per molecule is given.
An independent,  more local 
route to the insertion free energy will lead to strongly $\lambda$-dependent
results, thereby allowing a sensitive determination of $\lambda$
 from equating both 
expressions. The approach of thermodynamic integration introduced for RISM 
approaches  by Chandler \cite{chandl}
shall be used as it connects the pair correlation
functions on local distances to the thermodynamic properties. 
The variation of the free energy when turning on the interactions via the
Mayer $f$ function,  
$f_{\alpha\beta}^{(\zeta)} : \zeta \in [0,1] \mapsto 
f_{\alpha\beta}^{(\zeta)}$,
shall be used, where $f_{\alpha\beta}^{(1)}$ is the physical one and 
$f_{\alpha\beta}^{(0)}$ 
belongs to some known reference system; here the Greek indices run over the
colloid and the polymer segment sites. As only excluded volume interactions are
present, and because the limit of vanishing polymer segment size, 
$\sigma_p\to0$, is of interest, the thermodynamic integration is chosen to
take a mixture of polymers and colloidal point particles of free energy $F_0$ 
to the true mixture by growing the colloidal particles, $\sigma^{(\zeta)}_c=
\zeta \sigma_c$. From Ref. \cite{chandl} one then easily finds:
\begin{eqnarray}\label{m9}
\beta ( F - F_0 ) & = & 
\frac{\pi \varrho_c \varrho_p\sigma_c}{2} 
\int_0^1\!\!\!\! d\zeta (\sigma_p\!\!+\!\!\zeta\sigma_c)^2 
g^{(\zeta)}_{cp}(\frac{\sigma_p\!\!+\!\!\zeta\sigma_c}{2})  \nonumber\\ & + &
2 \pi \varrho_c^2\sigma^3_c 
\int_0^1\!\!\!\! d\zeta \zeta^2 
g^{ (\zeta)}_{cc}(\zeta\sigma_c)\; , 
\end{eqnarray}
where the $g^{(\zeta)}$ are the $\zeta$-dependent (e.g. via
the volume fraction of the colloid particles) pair correlation functions
which are evaluated at the distances of closest approach.
Equation (\ref{m9}) expresses that the growing colloidal spheres have to
push against the pressure of the surrounding system (polymers and colloids)
which --- in a virial theorem analogy ---
 is given by the probability of contact on the surface.
Immediately, one obtains a second independent result for the chemical potential
of \gl{m8} which, as argued, depends on $\lambda$ strongly as the packing
of polymer segments close to the colloidal particles, 
$g_{cp}(\frac12(\sigma_c+\sigma_p))$, enters crucially:
\begin{eqnarray}\label{m10} 
N \beta \delta \mu^{\rm (g)}_p|_{\varrho_p=0} & = & 
\frac{\pi \varrho_c\sigma_c N}{2}\; 
\int_0^1\!\!\!\! d\zeta (\sigma_p\!\!+\!\!\zeta\sigma_c)^2 g^{\rm
(\zeta)}_{cp}(\frac{\sigma_p\!\!+\!\!\zeta\sigma_c}{2})|_{\varrho_p=0}
  \nonumber\\ & + &
2 \pi \varrho_c^2\sigma_c^3 N\;
\int_0^1\!\!\!\! d\zeta \zeta^2 
\frac{\partial g^{\rm
(\zeta)}_{cc}(\zeta\sigma_c)}{\partial \varrho_p}|_{\varrho_p=0}\; . 
\end{eqnarray}
Even though it would be desirable to obtain $\lambda$ for all concentration
ranges, especially its form for low polymer concentrations is required.
On the one hand, even small amounts of polymers added to 
colloidal systems can strongly affect the phase diagram and colloid 
structure. On the other hand, if $\lambda$'s dependence on the
polymer parameters  is known then scaling considerations allow reasonable 
extrapolations of $\lambda$ into the semidilute region as will be shown in 
section V. Therefore, the two expressions, \gls{m8}{m10}, for vanishing polymer
concentrations will be used to obtain $\lambda$ in section IV.

\section{Solution in  low density limits}

The specified model of colloid-polymer mixtures covers all polymer and colloid
density regions, and distances  from the polymer repeat unit
size (ca. 5 \AA) up to the  collective correlation length approaching phase 
separation (some $\mu$m). Since small 
amounts of non-adsorbing polymer can alter the structure and phase diagram
of a colloidal system quite appreciably, 
simplification to 
consider rather low polymer concentrations is of initial interest. Under these
conditions there opens up a mesoscopic window where polymer segments are well
separated but polymer molecules  overlap and interact strongly. Such solutions
are called dilute or semi-dilute \cite{degennes}, and treatment of
this regime is most conveniently  done by performing the
  ``thread limit'' \cite{thread0,thread}, 
where  the size of a polymer segment  and the
corresponding statistical segment  size, $l_p/\sqrt{12}$,
 are taken to be negligibly small; 
$\sigma_p\propto l_p \to0$. In order to retain polymer molecules with a finite
 radius of gyration $R_g$,  the
number of repeat units is increased beyond bounds,
$N\to\infty$, such that  $\xi^2_0=l_p^2 N =
R_g^2/2=$ fixed. As has been shown by a rigorous solution of the
PRISM integral equations in \cite{threada}, intermolecular excluded volume
remains active if in parallel the monomer density is increased,
$\varrho_p\to\infty$, such that the number of polymer molecules per coil
volume ($\sim R_g^3$)
stays finite: $\varphi_p=2\pi \frac{\varrho_p}{N} \xi_0^3 
=$ fixed. The reduced polymer concentration $\varphi_p$
differs only by a numerical
factor from the often used polymer packing fraction $\eta_p
= \frac{4\pi}{3} \frac{\varrho_p}{N} R_g^3 = \varrho_p/\varrho_p^*
\approx\varphi_p/0.53$, where
$\varrho_p^*$, is the density when polymer coils start to 
interpenetrate. The mathematical thread
limit of the PRISM equations corresponds
to a scaling law description of the dilute-to-semidilute crossover of polymer
solutions and can be compared to field theoretic scaling laws and results
\cite{FuchsMueller}. In both cases only mesoscopic parameters, the
polymer molecule density and coil size, enter and all microscopic
parameters, like $\sigma_p$, $l_p$ and $\varrho_p\sigma_p^3$, drop out.
The effective polymer-polymer interaction becomes of the Edwards
delta-function 
type, $c_{pp}(r)=\hat{c}_{pp}(0) \delta({\bf r})$, where  the 
intermolecular excluded volume parameter, $\hat{c}_{pp}(0)$,
follows self-consistently from the
no-overlap condition, \gl{m2}. Besides its use for dilute and 
semi-dilute polymer solutions, 
experience also has shown when applying the
 thread limit outside its rigorous range of validity 
that it describes qualitatively adequately the spatially
 coarse-grained features of concentrated polymer solutions
and melts \cite{KSSMacro95}. 

Two limits of the scaling function of the
single chain form factor
are known in general and for Gaussian polymers simplify to:
$\omega(q=0) = N$ and $\omega(qR_g\gg1) \to (ql_p)^{-2}$. Note that the
self-scattering term, which is present in the full $\omega(q)$ 
of \gl{m4} for $q\sigma_p
={\cal O}(1)$, is not accessible in the thread limit.  In order to
keep simple and analytically tractable equations, the full intramolecular
structure factor shall be approximated by the standard
 Pad{\'e} interpolation between
the two asymptotes: 
\beq{d1}
\omega(q) 
\approx \frac{N}{1+q^2 \xi_0^2}\; .
\eeq
As the single-polymer structure factor is an input to our PRISM approach,
the use of random walk statistics in \gl{d1} for repelling coils
 can be considered  an additional
technical approximation in order to achieve analytical results. In order to 
capture effects of the non-trivial intramolecular correlations 
(``swelling'' and ``self-avoiding-walk statistics")
caused by intramolecular excluded volume, the equations (\ref{m1}) to 
(\ref{m10}) could be solved numerically with an appropriate $\omega(q)$ 
\cite{FuchsMueller}. Also, effects specific to semi-flexible polymers and
 arising from local chain rigidity are neglected in \gl{d1}, but
 could be incorporated into numerical studies. 

In the following (and on the rhs of \gl{d1}) 
dimensionless units shall be chosen by using the colloid diameter as unit
of length, $\sigma_c=1$. Then, the length scale ratio, 
$\xi_0=R_g/(\sqrt{2} \sigma_c)$, the 
relative polymer concentration, $\varphi_p$, and  the colloid packing
fraction, $\phi_c$,
 are the only remaining
physical parameters. Further notational simplification is provided by
defining $\bar{S}_{cc} = S_{cc}/\varrho_c$, 
$\bar{S}_{pp}= \frac{l_p^2}{\sigma_c^2 \varrho_p} S_{pp}$, $\bar{c}^s_{cp}
= c^s_{cp} \sigma_c^2/l_p^2$, and 
$\bar{\omega}(q)=  (l_p/\sigma_c)^2\omega(q)$.

The limit of considering only (semi-) dilute polymer solutions 
does not eliminate the non-linearities of the integral equations for the 
polymer and colloid structure. Insights into the physics described by the
m-PY PRISM equations and their  full solutions can be gained by reducing
one of the densities further to a dilute limit where at most pair-wise
direct
interactions of the diluted species can occur. This linearizes the
equations in the correlation functions of the diluted species and thus, 
as the
correlations of the majority component are known, simplifies the analysis.
These limits will be studied in the following, where in section III, the 
m-PY closure  parameter $\lambda$ still is kept arbitrary.

\subsection{Dilute Colloids}

In the limit $\phi_c\to0$ the equations simplify as the colloid particles 
do not alter the structure of the polymer fluid.  The collective
polymer structure factor 
for Gaussian intramolecular correlations within PRISM equals \cite{kcur3}:
\beq{dc1}
\hat{\bar{S}}_{pp}=\frac{\xi^2}{1+q^2\xi^2}\; ,
\eeq
where the polymer correlation length crosses over from $\approx R_g$
 at high dilution to the blob-size or density screening length for
concentrations within the semidilute regime: 
\beq{dc2}
\frac 1\xi = \frac{1}{\xi_0} + \frac{1}{\xi_\rho} = 
\frac{1+2\varphi_p}{\xi_0}\; .
\eeq
Note, that the neglect of microscopic length scales corresponds to the
assumption that the polymer concentrations in \gl{dc2} are far smaller than
melt densities, where segments of different polymer chains start to pack
densely (typically $\approx 30-40\%$ of melt density). 

The polymer segment profile  close to a single colloidal particle and the
resulting packing of two colloidal hard spheres is described by linear 
equations:
\begin{eqnarray}\label{dc3}
\hat{h}_{cp}(q) & = & \frac{\hat{\bar c}^s_{cp}(q)}{1+q^2\lam^2} 
\hat{\bar S}_{pp}(q)
\qquad\mbox{and}\qquad \nonumber\\
\hat{h}_{cc}(q) & = & \hat{c}_{cc}(q) + \frac{\varphi_p}{2\pi\xi_0}\; 
\frac{\hat{\bar c}^s_{cp}(q) }{
1+q^2\lam^2}  \hat{h}_{cp}(q) \; .
\end{eqnarray}
Closed equations for the pair correlation functions follow from 
and Eqs. (\ref{m2},\ref{dc3})
 and (\ref{m5}) -- (\ref{m7}). Even though these equations
can be solved by straightforward conversion to differential equations, 
a simplified Wiener-Hopf factorization of \gl{dc3} is used in Appendix A 
because it is close  to the solution technique of the full equations
for arbitrary $\phi_c$ and $\varphi_p$ to be published elsewhere, and 
can be presented in a more concise way. In Appendix A.1
it is shown that the functions $f_{ij}(r) = r ( g_{ij}(r) -1 )$ satisfy 
simple differential equations. The one describing the polymer segment 
density profile close to a single sphere is:
\beq{dc4}
(1+\lam \partial_r)(1+\xi\partial_r) f_{cp}(r) = 0 \fur r\ge \frac 12\; ,
\eeq
with initial conditions $f'_{cp}(\frac 12)= g'_{cp}(\frac 12)/2-1=-1$ 
and $f''_{cp}(\frac 12) =  g''_{cp}(\frac 12)/2 = 
(\frac 12 + \lam + \xi) /(\lam \xi)$, where a prime denotes a derivative. 
Also the direct correlation function at zero wave vector is obtained 
from \gls{ac2}{ac8}:
\beq{dc5}
\hat{c}_{cp}(0) =  \frac{-\pi\xi_0^2}{6N\xi^2}
(1 + 6 \xi + 12 \xi^2 + 6 \lam ( 1 + 2 \xi )^2 + 12 \lam^2 (1+2\xi) )\; .
\eeq
For dilute polymer solutions, $\varphi_p\to0$, the quantity $-N\, 
\hat{c}_{cp}(0)$
is the cross second virial coefficient describing the mutual excluded volume
between a hard sphere and a Gaussian polymer coil.

For the pair correlation function describing the 
probability of two isolated spheres in the polymer solution to be at a
separation $r$, the differential equation is:
\beq{dc6}
(1+\lam \partial_r) f_{cc}(r) + \frac{\varphi_p}{\xi_0} ( u + v \lam
\partial_r ) f_{cp}(r-\frac 12) = 0
 \fur r\ge 1\; ,
\eeq
with initial condition $f_{cc}(1)+1=g_{cc}(1)$.
The unknown parameters $u$ and $v$ follow from
 \gls{ac8}{ac12}, and the initial value is given by
the colloid pair contact value
at infinite dilution, $\phi_c\to0$: 
\begin{eqnarray}\label{dc7}
& g_{cc}(1) =
1 + & \nonumber\\ & \frac{\varphi_p\,\left( 2\,{\xi}^2 - 
       4\,{\lam}^3\,\left( 1 + 2\,\xi \right)  + 
       4\,\lam\,\xi\,\left( 1 + 2\,\xi \right)  + 
       {\lam}^2\,\left( 3 + 4\,\xi - 4\,{\xi}^2 \right)  \right) 
}{4\,{\xi}^2\,\xi_0} \; ,&
\end{eqnarray}
where corrections of  the order ${\cal O}(e^{-1/\lam})$ were neglected.
The contact probability for two colloids in principle can be calculated
exactly within m-PY, but the approximation to neglect corrections of 
${\cal O}(e^{-1/\lambda})$ simplifies the expressions greatly, and for
realistic values, (see below) where $\lambda <$ (mostly $\ll$)  0.31,
introduces errors of at most a few percent. 

From \gl{dc4} and the initial conditions follows that $g_{cp}(r)$ consists
of a superposition of two Yukawa tails, $A_\xi e^{-r/\xi}/r$ and
$A_\lam e^{-r/\lam}/r$, where $A_\xi>0$ and $A_\lam<0$ for $\lam<\xi$, and 
that  it describes a monotonous increase from zero to unity for 
$\frac 12 \le r < \infty$. Similarly, \gl{dc6} shows that $g_{cc}(r)$ is a
superposition of these two Yukawa tails (with constants $B_\xi$ and  
$B_\lam$) and of an additional secular term $C_\lam e^{-r/\lam}$.
 Nevertheless,
numerically it is found to decrease monotonously for all parameters.

Of interest is the second virial coefficient $B^c_2$ which follows from the 
(colloid partial) compressibility and measures the strength of the 
polymer induced pair potential \cite{hill}:
\beq{dc10} 
B^c_2 = \frac{-1}{2} \int d^3r\; h_{cc}(r)|_{\varrho_c=0} =
 B_2^{\rm HS} ( 1 - 3 \int_1^\infty
dr \; r f_{cc}(r)|_{\varrho_c=0} )\; ,
\eeq
where the result for hard spheres, $B_2^{\rm HS}= \frac{2\pi}{3}$, expresses
a purely steric repulsion, and negative values of $B_2$ indicate the presence
of a net attractive effective interaction due to ``depletion''.
 One finds from the solution of \gl{dc6}
\beqa{dc11}
 B^c_2/B_2^{\rm HS} &=& 1 -  \frac{3\,\varphi_p\,}{4\,{\xi}^2\,\xi_0} 
\nonumber\\ 
   &\times&  \left[ 2\,{\xi}^3\,\left( 1 + \xi \right)  - 
       4\,{\lam}^5\,
        \left( 1 + 2\,\xi \right)  + 
       {\lam}^4\,
        {\left( 1 + 2\,\xi \right) }^2 \right. \nonumber \\ &+& 
       4\,\lam\,{\xi}^2\,
        \left( 1 + 3\,\xi + 2\,{\xi}^2 \right)  + 
       4\,{\lam}^3\,
        \left( 1 + 4\,\xi + 6\,{\xi}^2 + 4\,{\xi}^3
          \right) \nonumber\\   &+& \left. 2\,{\lam}^2\,\xi\,
        \left( 3 + 11\,\xi + 12\,{\xi}^2 + 4\,{\xi}^3
          \right)  \right] \; , 
\eeqa
where again corrections of order ${\cal O}(e^{-1/\lam})$ were neglected but
could be evaluated straightforwardly. For $\lambda=0$, the PY result of Ref.
\cite{avik} is recovered.  The virial coefficient, as it  still contains 
the parameter  $\lambda$, will be discussed in section V, after the 
determination of $\lambda$ from thermodynamic consistency.

\subsection{Dilute Polymers}

In the limit of adding only a few  polymers to a dense fluid of hard spheres,
the m-PY PRISM equations simplify to:

\beqa{dp1} 
\hat{h}_{cp}(q) & = & \frac{\bar{\omega}_q 
\hat{\bar c}_{cp}^s(q)}{1+q^2\lam^2} \hat{\bar S}_{cc}(q)
\qquad  \mbox{and}  \nonumber\\ 
\hat{h}_{pp}(q) & = & \omega_q^2 \hat{c}_{pp}(q) 
+ \frac{6\phi_c}{\pi}\; 
\frac{\bar{\omega} \hat{\bar c}_{cp}^s(q)}{
1+q^2\lam^2}  \hat{h}_{cp}(q) \; ,
\eeqa
where the structure factor of the pure hard sphere fluid, $\hat{S}_{cc}$, is 
given in PY-approximation. The unperturbed  length scale characterizing
the polymer  correlations is given by $\xi_0$ as for an isolated polymer, but
colloid mediated interactions arise as the polymer coil is forced to squeeze
 into the voids between the spheres. In Appendix A.2 
it is shown that the functions $f_{ij}(r) = r ( g_{ij}(r) -1 )$ satisfy 
linear integro-differential equations with known integration kernels.
The average polymer segment density around  colloidal spheres
increases according to:
\beq{dp2}
(1+\lam \partial_r)(1+\xi_0\partial_r) f_{cp}(r) 
= 12 \phi_c \xi_0^2 \int_{-1/2}^{1/2}\!\! ds \; q_{cp}(s) f_{cc}(r-s)
\eeq
for $r\ge \frac 12$, where the initial conditions are:
$f'_{cp}(\frac 12)= g'_{cp}(\frac 12)/2-1=-1$ 
and 
\beq{dp3}
f''_{cp}(\frac 12) =  g''_{cp}(\frac 12)/2 = 
\frac{1-\phi_c+2(\lam+\xi_0)(1+2\phi_c)}{2\lam\xi_0(1-\phi_c)^2}\; .
\eeq
Here, the hard sphere function, $f_{cc}(r)$ is known or can be easily 
determined from \gl{dp1h} in Appendix A.2. 
The new factor function is given by:
\beq{dp4}
q_{cp}(r) = \frac a 2 ( r^2 - \frac 14 ) + b ( r - \frac 12 )  \; ,
\fur  -\frac 12 \le r \le \frac 12 \; , 
\eeq
and vanishes outside this range; the parameters are given in \gl{dp5}.
Again, the direct correlation function at zero wave vector, which is needed
for the thermodynamic calculation, can be obtained in explicit form
(see Appendix A.2):
\beqa{dp6} &
\frac{-6N}{\pi}\hat{c}_{cp}(q\!\!=\!\!0) = 
\frac{1}{1-\phi_c} & \nonumber\\ & + \frac{6( \xi_0 + \lam)}{(1-\phi_c)^2} +
\frac{12(\lam+\xi_0)^2 (1+2\phi_c)}{(1-\phi_c)^3} +
\frac{24\lam\xi_0(\lam+\xi_0) (1+2\phi_c)^2}{(1-\phi_c)^4}  \; .&
\eeqa
The microscopic colloid-segment hard core interaction causes a repulsive 
spike at contact which leads to an oscillatory behaviour  
for very large wave vectors:
\beq{dp7} 
\hat{\bar c}^s_{cp}(q\to\infty) = - 2 \pi
\cos{(\frac q2)}\; \frac{\lam}{\xi_0} \frac{1-\phi_c+2(\lam+\xi_0)(1+2\phi_c)}{
(1-\phi_c)^2} \; .  
\eeq
This remnant of the local segmental interactions also is present at
finite polymer concentrations, and for $\phi_c\to0$ is given by \gl{dp7}
with the replacement $\xi_0\to\xi$. 
The polymer pair correlation function follows from 
(\ref{ap6},\ref{ap8},\ref{ap10}) for $r>0$
\beqa{dp8} &
(1+\lam \partial_r)(1+\xi\partial_r) f_{pp}(r) = 
z_{\xi_0} e^{-r/\xi_0} + z_\lam e^{-r/\lam} & \nonumber \\ & +  
12 \phi_c \xi_0^2 \int_{-\frac 12}^{\frac 12} ds\; q_{cp}(s)
(r-s)(g_{cp}(|r-s|)-1) 
\; , &
\eeqa
where the initial conditions are $f_{pp}(0)=0$ and $f'_{pp}(0)=-1$, and
the parameters are given in \gl{ap9}. Clearly, by means of
the last term  the colloidal spheres imprint their local packing
structure onto the dilute polymers. In order to explicitly calculate the
pair correlation functions, numerical integrations of \gls{dp2}{dp8} are
most convenient using the known hard-sphere function,
 which can also be obtained from \gl{dp1h}.

Also of interest is the intermolecular excluded volume parameter which 
describes a colloid modified effective  interaction for polymer 
segments on different chains:
\beqa{ap11}
& \frac{\hat{c}_{pp}(0) \xi_0}{-8 \pi l_p^4}  =
\frac{b \xi_0^2+z_\lam+\bar{z}_{\xi_0}}{\lam + \xi_0} = 
\frac{1}{{\left( 1 - \phi \right) }^4\, \left( \lam + \xi_0 \right)}
  & \\ & \times \left[
    {\left( 1 - \phi \right) }^2\,
       \xi_0\,
       \left( 1 - \phi\,
          \left( 1 - 3\,\xi_0 \right) 
         \right) \right.  & \nonumber \\ & + 3\,{\lam}^2\,\phi\,
       {\left( 1 + 2\,\xi_0 - 
           \phi\,\left( 1 - 4\,\xi_0
              \right)  \right) }^2 
+      \lam\,\left( 1 - \phi \right) & \nonumber \\ &  \times \left.
       \left( 1 - \phi\,
          \left( 2 - 9\,\xi_0 - 
            12\,{\xi_0}^2 \right)   + 
         {\phi}^2\,\left( 1 - 9\,\xi_0 + 
            24\,{\xi_0}^2 \right)  \right) \right]
       \; .
& \eeqa
It simplifies to $\hat{c}_{pp}=-8\pi l_p^4/\xi_0$ for the pure polymer
system, and is connected to  the polymer second virial coefficient via:
\beq{ap13}
B_2^p = -\frac 12 \hat{h}_{pp}(q=0) = - \frac 12 N^2 \left(
\hat{c}_{pp}(0) + \frac{6\phi_c}{\pi} \; (\hat{c}_{cp}(0))^2 \; 
\hat{\bar S}_{cc} \right)\; ,
\eeq
which becomes 
$B_2^p = - \frac 12 N^2 \hat{c}_{pp}(0)\propto R_g^3$
 when no colloids are present.

\section{Enforcement of thermodynamic consistency}

The solutions for the structure of colloid-polymer mixtures open up two 
very different calculation routes to the thermodynamic properties. 
One, via the long wavelength fluctuations, termed ``compressibility 
route'', is suggested in \gl{m8} for the insertion free energy of polymers
into hard sphere fluids. The second, \gl{m10}, uses
very local information captured in the contact values of the pair correlation 
functions. Within integral equation theories for the structure of many-body
systems both results generally do
not coincide, and in the m-PY closure this aspect
of ``thermodynamic consistency'' is used to determine the effective 
interactions length scale $\lambda$ from equating both results.

The ideal gas result for adding point polymers, $\xi_0\to0$, immediately 
follows from \gls{m8}{dp6} and the observation
that $\lambda\propto \xi_0$ (corresponding to the polymer appearing as
an inpenetrable small sphere to the colliod) has to hold in this limit:
$N\beta\delta\mu^{\rm (c)}_p|_{\rho_p=0} = - \ln{(1-\phi_c)} + 
{\cal O}(\xi_0)$. This change in  
translational entropy when dissolving a point polymer in a sphere solvent 
of
packing fraction $\phi_c$ is connected to the free volume fraction
accessible to the polymer,  $e^{-N\beta\delta\mu_p} = V_f/V = 1 - \phi_c$.
The free volume for finite colloid but vanishing polymer
concentrations is a central quantity in the phase studies of the  free 
volume
approach by Lekkerkerker et al. \cite{Lekkerkerker92}.
In order to arrive at the same ideal gas limit from the 
polymer-colloid contact probability in \gl{m10}, since
$\frac{\partial}{\partial \varphi_p}  g_{cc}(1)|_{\rho_p=0} = 
{\cal O}(\xi_0^2)$ as follows from \gl{dc7}, 
 the following result has to hold for the contact
probability of polymer segments with the colloidal surface:
\beq{l1}
g_{cp}(\frac{1+\sigma_p}{2}) = \frac{1}{12N} \frac{1}{1-\phi_c} 
\fur \xi_0\to0\;.
\eeq
The contact value should be microscopically small, vanishing in the mesoscopic
limit, $N\to\infty$.
In the thread m-PY PRISM limit described in section II, where microscopic
parameters were scaled away, the colloid-polymer pair correlation function
agrees with this result.
Thread PRISM  correctly predicts on mesoscopic scales that 
$g_{cp}(\frac12)
\to 0$. In the case of (semi-) dilute polymer solutions a scaling 
connection 
of the segment-segment contact value to the mesoscopic scaling law 
description could be shown, which enabled one to estimate the 
microscopically small contact value  from an evaluation of the mesoscopic 
pair correlation function at a microscopic separation  \cite{FuchsMueller}.
From the parabolic polymer segment density profile close to a colloidal
particle and \gl{dp3}, one realizes that within m-PY this again is 
possible.
The mesoscopic pair correlation function (i.e. the thread $g_{cp}$)
 evaluated at a microscopic separation, 
$\tilde{\sigma}_p$, because of
$g_{cp}(\frac 12+\tilde{\sigma_p}/2) = \frac 12 g_{cp}''(\frac 12)\;
 \tilde{\sigma}_p^2$,
predicts the correct scaling behaviour of the contact value, \gl{l1}, 
\beq{l2}
g_{cp}(\frac{1+\sigma_p}{2}) = \frac{1}{2}\; \frac{\tilde{\sigma}_p^2}{
\lambda\xi_0}\; \frac{1}{1-\phi_c} \fur \xi_0\to0\; ,
\eeq
and thus recovers the ideal gas result for the chemical potential from the
``wall virial'' route. Evidently, however, the microscopic 
distance could only be calculated if the full PRISM equations also 
including
microscopic length scales were solved. As this has not been achieved yet,
in Ref. \cite{Fuchs00} for simplicity   the distance 
$\tilde{\sigma}^2_p=2 l_p^2$ was chosen, because
then the distance over which point like polymers rearrange close to a colloid
sphere becomes numerically identical to the polymer correlation length:
$\lambda=\xi_0$ for $\xi_0\to0$. Note that this holds for all colloid
packing fractions, and that  $\tilde{\sigma}_p$  has to be chosen only
once.

It is the major difference of the novel m-PY closure to the previously studied
PY description, where $\lambda=0$, and (as can be expected)
 to prior numerical solutions
of the PRISM-PY equations \cite{Khalatur,Ferreira00}, 
that thermodynamic consistency in
$\delta \mu_p$ is possible in a scaling sense.  For $\lambda=0$ in the 
PY-case, the mesoscopic pair correlation function predicts a scaling of the 
colloid-polymer contact value as $g_{cp}(\frac{1+\sigma_p}{2}) =
\tilde{\sigma}_p/(\xi_0(1-\phi_c))$ for $\xi_0\to0$,
thus violating \gl{l1} and the ideal gas insertion free energy. 
Obviously, the PY closure overestimates the segment density of polymers, 
even of an isolated polymer, close to walls or colloidal spheres, as a 
factor $\sqrt N$ too high contact probability is predicted. This comparison
justifies the motivation and the physical interpretation of the m-PY closure 
in section II.
Additionally, the PY-approximation $\lambda=0$ predicts
a segment density profile which increases much more strongly,
$g^{\rm PY}_{cp}(r) = 
[\frac{1}{\xi_0(1-\phi_c)} + \frac{2+4\phi_c}{(1-\phi_c)^2} ]
(r-\frac12) + {\cal O}((r-\frac12)^2)$, compared to
$g_{cp}(r) = \frac 12 g''_{cp}(\frac12)\; (r-\frac12)^2
 + {\cal O}((r-\frac12)^3)$ from \gl{dp3}, 
 and thus strongly
underestimates the polymer induced depletion attraction. 
This follows since for small
separation of two colloidal particles, PY predicts a linear increase of the
polymer density in the gap, whereas the correct
\cite{eisenriegler,joanny} and m-PY result for Gaussian
polymers is a quadratic increase.

\begin{figure}[h]
\centerline{{\epsfysize=9.cm 
  \epsffile{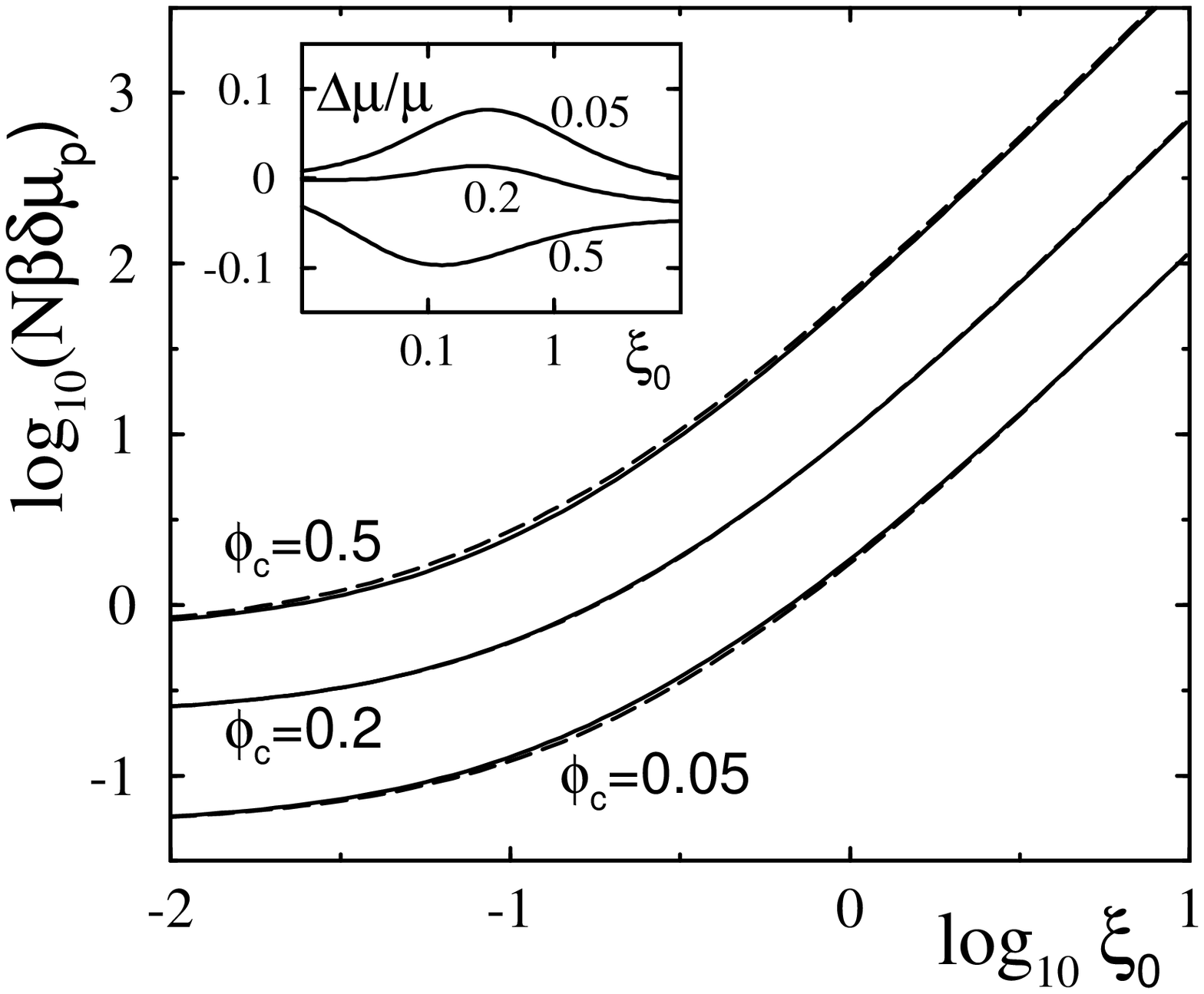}}}
\caption{Polymer molecule excess chemical potentials, 
$\delta\mu_p|_{\varrho_p=0}$, as functions of the size ratio $\xi_0$
from the two different routes used to enforce thermodynamic consistency,
for three colloid packing fractions as labelled. The solid lines give the
results from  long wavelengths, Eq. (\protect\ref{m8}), the dashed lines the
corresponding
 ones from local packing, Eq. (\protect\ref{m10}).  The inset shows
the relative errors, which for all parameters are smaller than 15 \%.
\label{fig3}}
\end{figure}

Extension of the expansion for small $\xi_0$ leads to the higher order result
$\lambda = \xi_0 - 4 \xi_0^2 + \xi_0^3 
\tilde{\Lambda}(\phi_c)+\ldots$, where the 
dependence on the sphere fluid packing density indicates an anomalous behaviour
of $\lambda$ for $\phi_c\to1$; i.e. one finds $\tilde{\Lambda}(0)=16$ but
$\tilde{\Lambda}(\phi_c)
\to - \frac{144}{(1-\phi_c)^2}$ for $\phi_c\to1$. As $\lambda$ is restricted to
positive values, a different expansion scheme is required for high colloid
packing fractions, and for large polymers or small colloidal particles.
It is described in  Appendix B. The calculation leads to 
$\lambda \to \Lambda(\phi_c)$ for $\xi_0\to\infty$ where:
\beq{l5} 
\Lambda(\phi_c) \to \left\{ \begin{array}{ll}
\frac{1}{\sqrt5+1} - 0.80\ldots  \phi_c + \ldots & \phi_c\to0 \\
 & \\
\frac{1/3}{\sqrt{5}+1}(1-\phi_c) + 0.78\ldots  (1-\phi_c)^2 + \ldots 
& 
\phi_c\to 1 
\end{array}\right.\; ,
\eeq
which verifies the interpretation of $\lambda$ as the range of polymer
segment rearrangement, because it is always less than
the smaller of the two molecular sizes, and becomes vanishingly small if no
free volume is available for the polymers. Exact determination 
of $\lambda$ 
from \gls{m8}{m10} for all parameter values is difficult, because 
thermodynamic integrations are required. 
Thus we proposed the
approximation to interpolate  between the known exact
limits \cite{Fuchs00}:
\beq{r1}
\lambda^{-1}= \xi^{-1} +
\frac{1+2\phi_c}{1-\phi_c}\frac{\lambda_1}{\sigma_c}\; ,
\eeq
where $\lambda_1=(\sqrt{5}+1)$, $\xi=\xi_0$ for vanishing polymer 
concentration, and for convenience dimensional units are restored.
This Pad{\'e} approximation
satisfies the thermodynamic consistency condition  from \gls{m8}{m10}
up to relative errors of 15 \% for all parameter values, see Fig. 
\ref{fig3}. For small polymers, $\lambda$ follows the polymer 
correlation length. For large single
 polymer coils, $\lambda$ is expected to be determined by the pure
 hard sphere fluid correlation length beyond which density
 fluctuations are screened, which is $\phi_c$-dependent. The result
 in \gl{r1} can be shown to be in excellent agreement with this
 intuitive idea.

The expression in \gl{r1} immediately suggests an extrapolation for 
$\lambda$
from its present calculation at vanishing polymer concentration 
to finite $\varphi_p$. If the  polymer correlation length of the
dilute situation, $\xi_0$, is replaced with the full density dependent 
one, 
$\xi$ from \gl{dc2}, then $\lambda$ also is known for $\phi_c=0$ but now
arbitrary polymer concentrations,  $\varphi_p>0$. This replacement is 
suggested
as the role of the single polymer molecule structure factor, $\omega(q)$,
 in \gl{dp1} is taken over by the collective one, $S_{pp}(q)$, if the
densities are changed accordingly.  This standard procedure from
polymer-scaling approaches thus substitutes the blob-size or 
density screening length in place of the chain size for semi-dilute
situations \cite{degennes} 
  In section VI, it will be shown
that this extrapolation achieves thermodynamic consistency 
for the 
insertion free energy of adding colloids to a polymer solution, i.e. for
 quite
a different thermodynamic quantity as originally considered for the
determination of $\lambda$.  
In Ref. \cite{Fuchs00}, this way of extrapolating $\lambda$ to
 finite polymer
and colloid
concentrations was suggested and used with 
$\xi=\xi(\phi_c,\varphi_p,\xi_0)$ being the full polymer 
correlation length, which depends on  $\phi_c$, $\varphi_p$ and $\xi_0$. 
In the
two cases considered here it simplifies to 
$\xi(\phi_c,\varphi_p=0,\xi_0)=\xi_0$ and
$\xi(\phi_c=0,\varphi_p,\xi_0)=\xi_0/(1+2\varphi_p)$, see \gl{dc2}.

\section{Results for colloids diluted in a polymer solvent}

If only a small amount of colloidal hard spheres is dissolved in 
a polymer solution, then the structure of the polymer fluid is not 
affected. 
The intramolecular density fluctuations or the form factors for the 
individual
polymers, $\omega(q)$, are rather well understood from field theoretic
considerations \cite{schaf99}. PRISM describes the packing, i.e. the 
intermolecular correlations, of the polymeric macromolecules which are 
simplified to Gaussian chain molecules in the present work. The
 intermolecular
pair correlation function exhibits the well-known ``correlation hole'',
which shows that polymer molecules for entropic reasons softly repel
each other. The correlation hole has a non-trivial structure on the 
length 
scale of the size of the molecule (radius of gyration) and also for shorter
distances, on the mesh- or blob scale characterised by the density 
screening
length $\xi$. 

\begin{figure}[h]
\centerline{{\epsfysize=9.cm 
  \epsffile{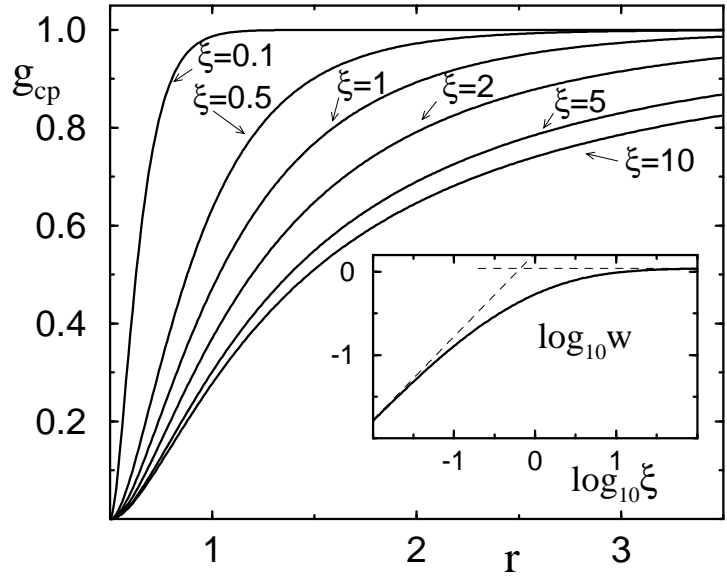}}}
\caption{Polymer-colloid pair correlation function, $g_{cp}(r)$, exhibiting
 the polymer segment depletion layer close to an isolated colloidal sphere for
various polymer correlation lengths $\xi$, as labelled.
The inset shows a double logarithmic plot of the width $w$
of the depletion layer, defined by $g_{cp}(r=\frac 12 + w)=\frac 12$, as a
function of $\xi$; thin dashed lines mark
the asymptotes $w\to1.66 \xi$ ($w\to 1.1$) for
$\xi\to0$ ($\xi\to\infty$) respectively. 
\label{fig4}}
\end{figure}

Close to the diluted colloidal particle the polymer segment density is 
less than in bulk
 resulting in a depletion layer of varying width.
Its form is described by \gl{dc4} and shown in Fig. \ref{fig4}. The width 
is  
given by the polymer correlation length $\xi$ as long as the particle
is much larger. Only for dilute polymer solutions, this correlation length
agrees with the polymer size, $\xi_0$. In the semidilute region, it is 
given 
by the blob size or mesh width. If the particle becomes smaller than 
the polymer correlation length, then the depletion layer width crosses over
to 
 the particle diameter. This is shown  in 
the inset in Fig. \ref{fig4}. If very small particles are immersed in the
polymer solution, then the depletion layer has an additional
 power law tail due to chain connectivity correlations which becomes
$g_{cp}(r\gg1) = 1 - (\frac 12 + \frac{1}{\lambda_1} )\; 
\frac{\sigma_c}{r}$ for $\xi\to
\infty$ \cite{eisenriegler}. 

 In the dilute polymer solution limit, the PRISM result
compares rather well with field theoretic  calculations from Ref.
\cite{eisenriegler}. While the width is of the same order  in both results,
especially the small distance power-law increase of the density profile,
$g_{cp}(r)=\frac12 g_{cp}''(\frac{1}{2})(r-\frac{\sigma_c}{2})^2
+\ldots$, 
can be compared as it obeys a scaling law, where the (universal) amplitude 
of PRISM (semi-) quantitatively agrees with the known field theoretic (FT)
limits:
\beq{rc1}
g_{cp}''(\frac12) \to \left\{\begin{array}{lccccr}
\frac{1}{\xi_0^2} & {\rm (PRISM)} & \quad & \frac{1}{\xi_0^2} & {\rm (FT)} &
\xi_0\to0 \\
 & & & & \\
\frac{2\lambda_1}{\sigma_c^2} & {\rm (PRISM)} &  \quad &
\frac{8}{\sigma_c^2} & {\rm (FT)} & \xi_0\to\infty 
\end{array}\right. \; .
\eeq
The 
PRISM result for the polymer profile in \gl{dc4} extends into the 
semi-dilute concentration region where  mean field calculations 
for the case of $R_g\ll \sigma_c$ are available \cite{joanny}. Again, the
amplitude, $g_{cp}''(\frac12) \to  \frac{1}{\xi^2}$ for $\xi\to0$, 
agrees verifying its  universality \cite{eisenriegler,eisenriegler2}. Only in 
the case of very large polymers, m-PY overestimates the depletion effect
by ca. 20\% in \gl{rc1}.
  Figure \ref{fig4}
shows that the width of the depletion layer for semidilute solutions is 
set by
the polymer correlation length (for $\xi_0\ll\sigma_c$), 
and that the molecular size, $R_g$ plays
no role for the depletion layer; this also is evident from \gl{dc4} where
only the correlation length of the collective polymer fluctuations appear.
Using an effective pair potential approach
\cite{Asakura54,Gast83,Lekkerkerker92} 
determined under dilute polymer conditions
strongly overestimates the range of the depletion 
interactions in semidilute polymer solutions.

\begin{figure}[h]
\centerline{{\epsfysize=9.cm 
  \epsffile{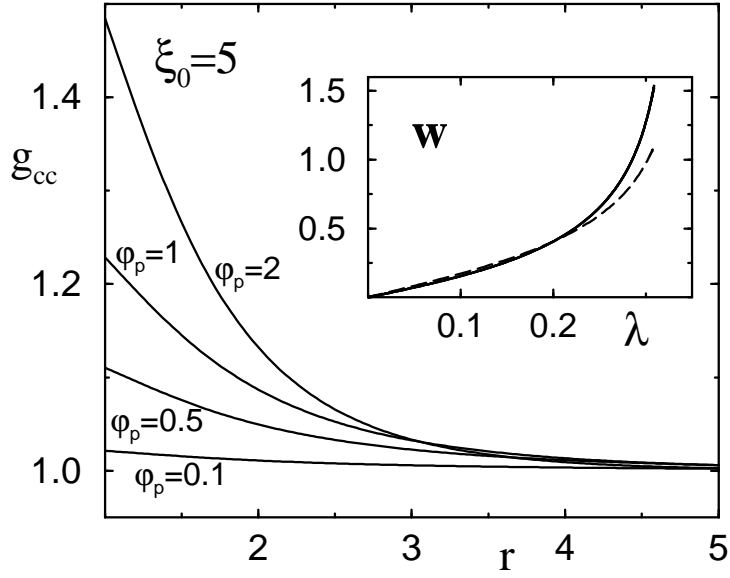}}}
\caption{Colloid pair correlation function, $g_{cc}(r)$, 
for two isolated hard spheres in a solution of polymers with size
ratio $\xi_0=5$ for the denoted polymer concentration $\varphi_p$.
The inset shows as a solid line
the range $w$ of the depletion attraction estimated by
$g_{cc}(r=1+w)=\frac 12(1+g_{cc}(1))$ as a function of the colloid-polymer
interaction length $\lambda$. Curves for various polymer concentrations
 ($\varphi_p=$ 0.1, 1, 2 ,10) overlap, while the size ratio  $\xi_0$ runs
between $0.01 \le \xi_0\le 100$. The dashed line compares the width of the 
depletion layer in $g_{cp}(r)$ from Fig. \protect\ref{fig4}.  
\label{fig5}}
\end{figure}

The polymer induced depletion attraction becomes apparent from the colloid
pair correlation function describing the probability of two isolated colloidal
spheres to be at a distance $r$, \gls{dc6}{dc7}.
As seen in Fig. \ref{fig5}, for  close distances  and especially at
contact, this probability is increased above the random value of
unity.  The $g_{cc}(r)$ decrease from the contact values,
\gl{dc7}, monotonously without 
oscillatory features or layering. Thus, the effective or induced 
potential, $\beta V^{\rm eff.}(r) = -\ln{g_{cc}(r)}$, is attractive and 
monotonic, and does not exhibit repulsive barriers. 
The contact value scales like $g_{cc}(1)-1\sim \frac{\varphi_p}{\xi_0}$ in
either limit of very small or very large polymers, i.e. $\xi_0\to0$ ($\infty$).
The distance characterising 
the decay of $g_{cc}(r)$ is closely connected to the width of the depletion
layer as seen in the inset of Fig. \ref{fig5}, and both depend on the
polymer parameters only via the nonlocality length $\lambda$. The 
correlation between
two colloidal particles which are much smaller than the polymer mesh width
falls off for large distances as $g_{cc}(r\gg1) \to 1 + (1+2/\lambda_1)^2 
\frac{\pi l_p^2\sigma_c^2\varrho_p}{r}$ for $\xi\to\infty$, 
indicating a weak but 
long-ranged attraction
\cite{eisenriegler3}. It becomes screened at the polymer or
blob size only.

\begin{figure}[h]
\centerline{{\epsfysize=9.cm 
  \epsffile{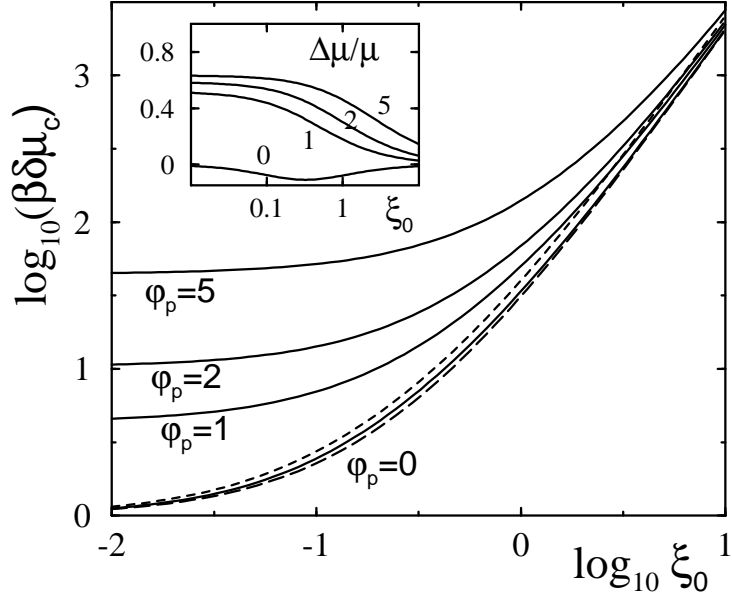}}}
\caption{Colloid excess chemical potentials at infinite dilution,
$\beta \delta\mu_c|_{\varrho_c=0}$,
as functions of the polymer to colloid size ratio $\xi_0$
for four polymer densities as labelled. The solid lines give the
results from long wavelengths,
 Eq. (\protect\ref{e3}), the dashed line is the result
 from local packing, Eqs. (\protect\ref{e4},\protect\ref{r2}), 
in the dilute polymer limit. The 
short dashed line compares the field theoretic result known in
 the dilute limit
 \protect\cite{eisenriegler}. The inset shows the relative errors in 
thermodynamic consistency for the four polymer densities.
\label{fig6}}
\end{figure}

A measure of the tendency of colloidal spheres to dissolve in polymeric fluids
is given by the chemical potentials or solvation energies. From  the
 compressibility route in analogy to \gl{m8}, one finds:
\beq{e3}
\beta \delta\mu^{\rm (c)}_c|_{\varrho_c=0} = - \int_0^{\varrho_p} d\varrho'_p
\hat{c}_{cp}(q=0,\varrho_p')|_{\varrho_c=0}\; .
\eeq
From the free energy (see \gl{m9}) required to grow the colloid particle 
from a point to its actual size one finds:
\beq{e4}
\beta \delta \mu^{\rm (g)}_c|_{\varrho_c=0} = 
 \frac{\pi c_p\sigma_c^3\xi_0^2}{2} \int_0^1 d\zeta\, \zeta^2 
g^{\rm (\zeta)}_{cp}\ \!\!''(\frac\zeta2)|_{\varrho_c=0}
\; .
\eeq
For dilute polymer solutions, where $\xi=\xi_0$, the result from both
routes obeys thermodynamic consistency (up to errors of 10 \%)
and agrees with field theoretic results up to 20 \%
 as shown in fig. \ref{fig6}.
From the local packing information, \gl{e4}, an explicit result can be found:
\begin{eqnarray}\label{r2}
\beta \delta \mu^{\rm (g)}_c|_{\varrho_c=0} = \frac{\pi c_p \sigma_c^3
\xi_0^2}{6\xi^2} 
(1 + (6 + \frac 32\lambda_1) (\frac{\xi}{\sigma_c}) + 
6\lambda_1 (\frac{\xi}{\sigma_c})^2 ) \; ,
\end{eqnarray}
which has a number of polymer specific features.
For a fluid of small polymers or far into the semi-dilute regime,
 $\xi\to0$,
inserting a colloidal sphere costs the free energy of creating its volume 
by doing work against the
osmotic pressure of the polymer fluid. For larger polymers, the form of the
polymeric coil enters, and in the limit of very large polymer chains the 
chemical potential becomes independent of $R_g$ and
scales linearly with the length over which polymer segments need to be 
rearranged, which is just the colloid diameter $\sigma_c$. As the added
sphere sees local strands of the polymer network only, the result for small 
spherical
colloids becomes independent of the polymer's size ($R_g$) or degree of 
polymerization, and also independent of the mesh or blob size.
Thus it becomes independent of the polymer molecule concentration
\cite{Eisenriegler00b}:
\beq{e5}
\beta \delta \mu_c|_{\varrho_c=0} \to \pi \rho_p \sigma_c l_p^2
 \lambda_1 \fur 
\xi\to\infty \; ,
\eeq
which holds in the dilute and semi-dilute region.
This result also quantitatively compares favourably to the known
behaviour in the dilute limit \cite{eisenriegler},
$\beta \delta \mu_c|_{\varrho_c=0} \to 4 \pi \rho_p \sigma_c l_p^2$.
For the semidilute region, the two routes in \gls{e3}{e4} predict
$\beta \delta\mu^{\rm (c)}_c|_{\varrho_c=0} \to
\frac13\; \beta \delta \mu^{\rm (g)}_c|_{\varrho_c=0}$ and
\beq{e6}
\beta \delta\mu^{\rm (c)}_c|_{\varrho_c=0} \to
\frac{\pi}{6} \sigma_c^3 \beta \Pi \;\;,\;\;
\beta \Pi = 
\frac{4}{3} c_p  \varphi_p^2 \; , \fur \varphi_p\to\infty \; ,
\eeq
which explicitly states the connection to the osmotic pressure, $\Pi$,
 of the pure
polymer system, which is given in PRISM approximation for Gaussian 
polymers 
in \gl{e6}.
Note, that the extrapolation of $\lambda$ to finite polymer concentration
leads to thermodynamic consistency also in these cases (as claimed at
the end of section IV), as both routes, 
\gls{e3}{e4}, lead to \gl{e5} identically, and the results in \gl{e6} only 
differ by a numerical prefactor, which could be absorbed into a 
density dependent microscopic length, $\tilde{\sigma}_p(\varphi_p)$.

\begin{figure}[h]
\centerline{{\epsfysize=9.cm 
  \epsffile{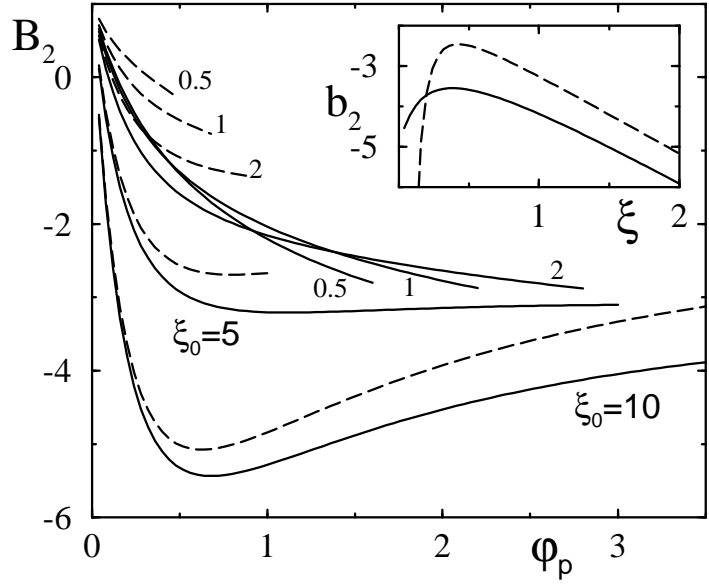}}}
\caption{Colloid second virial coefficient $B_2^c$ in units of the
hard sphere result, $B_2^{\rm HS}=\frac{2\pi}{3}$, as a function of the
polymer concentration for various size ratios $\xi_0$ as labelled.
The solid lines present the
results from long wavelengths, 
 Eqs. (\protect\ref{dc11},\protect\ref{e7},\protect\ref{e8}),
 the dashed lines present the results
 from local packing, Eq. (\protect\ref{e10}). The curves are cut in order to 
prevent overcrowding the figure.
The inset shows the semidilute scaling law, $\tilde{b}^c_2(\xi)$
from Eq. (\protect\ref{e9}), which 
applies to situations where the polymer correlation length $\xi$ (blob 
 or mesh size) is of the order of the colloid diameter.
The solid line again follows from Eq.
 (\protect\ref{dc11}) and the dashed one from
Eq. (\protect\ref{e10}).
\label{fig7}}
\end{figure}

The mean polymer induced depletion attraction for two colloidal spheres can be
quantified using the second virial coefficient $B^c_2$ which  is 
accessible experimentally. Representative results are shown in Fig \ref{fig7}.
The m-PY result, \gl{dc10}, expresses that  the effective
colloidal pair potential depends on the polymer concentration and the 
polymer
to colloid size ratio.
For small polymers, \gl{dc10} simplifies to:
\beq{e7}
\frac{B_2^c}{B_2^{\rm HS}} \to 1 - \frac{12\varphi_p}{1+2\varphi_p} +
\frac{3\,\left(28\,\lambda_1 - 53 \right) \,
    \varphi_p\,\xi_0}{4\,
    {\left( 1 + 2\,\varphi_p \right) }^2 \sigma_c} + \ldots
\;\mbox{for }\; \xi_0\to0 \;,
\eeq
which indicates an appreciable attraction between the two colloids. 
Note that the third term is positive and
 predicts a weakening of the attraction when
the polymer size starts to grow. Increasing
the polymer density strengthens the attraction which saturates at a
finite negative value.
The reliability of these results is discussed at the end of this 
section and in appendix C where alternative closures are examined.

For very large polymers, 
again an effective two-colloid attraction is observed.
\beqa{e8}
\frac{B_2^c}{B_2^{\rm HS}}  \to  1
& - &  \frac{3(2+\lambda_1)^2}{2\lambda_1^2}
\frac{\varphi_p(\xi_0/\sigma_c)}{(1+2\varphi_p)^2}  \nonumber\\ & - &
\frac{( 24 + 
    18\,\lambda_1 + 3    \lambda_1^2\, ) \varphi_p } {2\,
    \lambda_1^2\,
    \left( 1 + 2\,\varphi_p \right)}
\fur \xi_0\to\infty \;.
\eeqa
As seen in Fig. \ref{fig7},
 two hard spheres immersed in a fluid of much larger 
polymers feel an induced attraction which varies non-monotonically
with 
polymer concentration and increases with polymer size as previously
found based on the $\lambda=0$ PY closure \cite{avik}. Its origin is
purely entropic as the added particles hinder conformational fluctuations
of a polymer strand they are embedded in. For large distances, the
conformational entropy loss at both particles 
is additive. On shorter distances, however,
chain connectivity restricts the conformational fluctuations of a polymer
molecule even without added particles. Thus the additional
 entropy loss is smaller
if the two particles are close. This induces  an effective long-ranged
attraction among the colloidal particles.
For small concentrations, adding polymer strengthens the
effective attraction. Around the dilute-semidilute crossover density,
however, the effective  
range of the polymer induced attraction  crosses over from the polymer
size to the mesh size, and thus starts to decrease appreciably. The latter
effect dominates in the semidilute concentration region and therefore
$B_2^c$ decreases again. 
It is important to mention that a mean field approach \cite{schaf99}
 employing a RPA 
 approximation, where $\xi\sim 1/\sqrt{\varrho_p}$,
 would be too crude to handle this
competition and would miss  the increase of $B_2^c$ above the
overlap concentration. Apparently, a simple superposition approximation
which decouples the depletion layers 
around each of the colloid spheres misses the long-ranged, but weak, attraction
\cite{Tuinier00}. The minimum value of $B_2^c$ for the considered Gaussian
polymer statistics is  
$-\frac{3(2+\lambda_1)^2}{16\lambda_1^2} (\xi_0/\sigma_c)\approx -0.17 
(R_g/R_c)$,  which is asymptotically deeper
 than the value $-0.50 (R_g/R_c)^{0.40}$ for self-avoiding walk 
polymer statistics \cite{Eisenriegler00b}. The more swollen polymer 
molecules in good solvents apparently are more open to the particles and allow
stronger  interpenetration so that the induced colloid attraction 
is smaller \cite{eisenriegler3,eisenrieglervergleich}.

Deep in the semidilute density region, $\varphi_p\to\infty$, 
the second virial
coefficient again saturates, $B_2^c/B_2^{\rm HS}\to-5$, indicating a finite
effect of the pair-wise attraction which is independent of the size ratio.
This limit corresponds to a vanishingly small polymer blob size.
 Yet, for large polymers there can open up a window, 
where adding polymer only slightly
 changes $B_2^c$, in a manner which 
is much weaker than at the overlap concentration
but still differs from the asymptotic value. This is the polymer concentration
range where the polymer mesh width is not yet negligible compared to the
particle size, and there  a broad maximum develops:
\beq{e9}
\frac{B_2^c}{B_2^{\rm HS}} \to \tilde{b}_2^c(\xi)
\fur \varphi_p \to\infty\; ,\; \xi_0\to\infty
\quad\mbox{with}\quad \xi={\rm const.}\;.
\eeq
The scaling function $\tilde{b}^c_2$ is shown in the inset of Fig. \ref{fig7}.
For semidilute polymer concentrations, the depletion attraction among two 
colloidal spheres depends non-monotonously on the ratio of the blob to
the sphere size. There is an optimal blob size which roughly equals the
sphere radius, where the induced attraction is minimal, or $B_2^c$
maximal. This makes physical sense since for $\sigma_c/2\approx\xi$
the particles can ``just fit in'' the polymer mesh spaces without
distorting it.
For larger polymer correlation lengths, the crossover to \gl{e8} sets in 
because the range of attraction increases. Unexpectedly, however, also for
smaller $\xi$,  $B_2^c$ becomes more negative, presumably because the depth
of the attraction increases.

There does exist a caveat for the above results on the 
colloid-colloid
interactions \cite{avik,Dijkstra99,Fuchs00,Schmidt00}.  Liquid state
theory (and also \cite{Gast83})  with PY-closure,
 density functional approaches to the colloid structure, and free
 energy based approaches \cite{Lekkerkerker92}, all underestimate
the depletion attraction in situations where it far exceeds $k_BT$. This
error, which presumably affects the results for dilute colloidal 
particles in
a dense solvent of small polymers, $\phi_c\to0$, $\xi_0\ll\sigma_c$ and
$\varphi_p\gg1$, 
arises from an inherent linearization of the depletion potential
in the considered binary-mixture approach. Therefore it is fortunate
 that a
thermodynamic consistency condition can be formulated explicitly 
addressing the
accuracy with which the depletion attraction is handled.
 There exists another  independent  expression for 
 $B^c_2$ which follows from \gls{m9}{dc3} and the definition 
\gl{dc10}:
\beqa{e10} &
\frac{B^c_2}{B_2^{\rm HS}} =  3 \left[ \int_0^1d\zeta\, \zeta^2 
g^{(\zeta)}_{cc}(\zeta) +
\frac{\varphi_p}{48\xi_0}\;
\int_0^1d\zeta\, \zeta^5 
  \frac{\partial g^{(\zeta)}_{cp}\ \!\!''(\zeta/2) }{\partial \phi_c}
\right. & \nonumber \\ & \left.
- \frac{\varphi_p\xi^2}{32 \xi_0} \left(\int_0^1d\zeta \, \zeta^2  
g^{(\zeta)}_{cp}\ \!\!''(\zeta/2)\right)^2 \right]_{\rho_c=0} \; . &
\eeqa
This result follows from the local packing information and
 shows a very different, and unphysical, scaling compared with the result 
from large wave length fluctuations  for small polymers and/or high polymer 
concentrations:
$\frac{B^{c {\rm (g)}}_2}{B_2^{\rm HS}} \to \frac{1}{24}
 (\varphi_p/\xi_0)^3$.
This throws severe doubts on the result in \gl{e7} and on our
treatment of the depletion attraction in these two cases, $\xi_0\to0$ or
$\varphi_p\to\infty$, where from a pair potential point of view the 
depletion 
attraction should be arbitrarily large. Thus m-PY PRISM with the PY
closure for colloid-colloid correlations apparently
cannot capture the strong induced attraction in these two limits of
theoretical interest as has been pointed out previously 
\cite{avik,Fuchs00}. As the second 
virial coefficient can be considered a worst
case example of this failure, appendix C examines this issue using different
theoretical approaches.   
 Reassuringly, the results for the colloidal second 
virial coefficient for large polymers are recovered semi-quantitatively
from the two routes within m-PY PRISM.
The scaling in the limit of very large polymers is
recovered exactly, i.e. the term diverging linearly with $\xi_0$ in \gl{e8}.
Equation (\ref{e10}) also predicts  that a scaling law exists for the 
second virial coefficient of colloidal particles in a polymer mesh  which 
semiquantitatively compares with the one of \gl{e9}, see Fig. \ref{fig7}.
Intriguingly it also shows the non-montonic behaviour of 
$\tilde{B}_2^c(\xi)$
for intermediate $\xi$.

\section{Results for polymers diluted in a hard sphere fluid}

While numerous field theoretic results for polymer solutions exist,
including for the question of dissolving one or two particles in dilute 
solutions,  
and could  serve for tests of the m-PY results in the previous 
section, to our knowledge little is
known about the packing of dilute polymer chains into dense particle 
fluids. The case of present interest is when the 
 particles are much larger than  the repeat
units of the polymer chain, and both components are immersed 
in a small molecule background solvent treated as a continuum. 
If the amount of added non-absorbing 
polymer is small,
then the structure of the colloidal fluid is not changed. In the present 
treatment it is given by the reliable and easily improved PY-theory 
description. 

\begin{figure}[h]
\centerline{{\epsfysize=9.cm 
  \epsffile{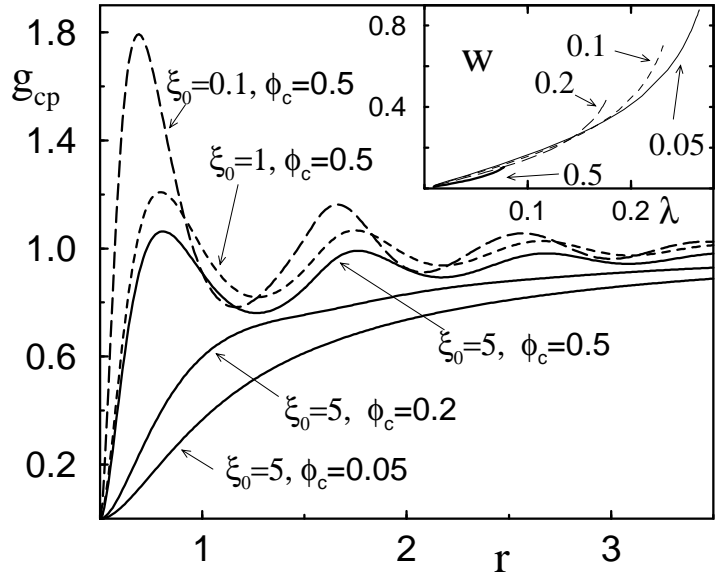}}}
\caption{Polymer-colloid pair correlation function, $g_{cp}(r)$, 
for a polymer added to a hard sphere solution; colloid packing
fraction $\phi_c$ and size ratio $\xi_0$ as labelled.
The inset shows the width $w$ of the depletion layer 
(defined as $g_{cp}(r=\frac 12+w)=\frac 12$) versus the interaction range
$\lambda$ for the labelled colloid packing fractions $\phi_c$. The curves 
are parametrized by the size ratio $\xi_0$ which varies as $0.01 \le \xi_0
\le 100 $ with increasing $w$ and $\lambda$.
\label{fig8}}
\end{figure}

Whereas the segmental depletion layer of a single polymer (or a semidilute 
solution) around a single (spherical) particle exhibits a monotonic 
dependence  on the distance to the surface of the particle,  adding further
particles forces the polymer to squeeze into the open spaces. Thus
the probability of finding polymer segments at a distance $r$ from the
center of a colloidal particle,  $g_{cp}(r)$,
develops an oscillatory structure whose period
is correlated with the colloidal size. In Fig. \ref{fig8} the evolving 
layering
of the segment density is shown. For high particle densities, the polymers
pack tightly 
 into the voids and thus are  close to the particles. The depletion
layer, even though present in principle, is restricted by the colloid spacing,
and varies with external parameters.
Thus, the assumption of an effective pair potential 
becomes inappropriate, because it requires that the range of
the polymer induced colloid-colloid potential, which naturally 
is connected to
the depletion layer, itself becomes particle density dependent. 
The inset of Fig. \ref{fig8} shows that there is a strong correlation 
of the
colloid-polymer interaction range, $\lambda$, to the width in the 
depletion
layer. While varying the polymer size by 4 orders of magnitude,
 neither the width nor 
$\lambda$ change as strongly. Moreover, for somewhat higher colloid solvent 
concentrations
the width becomes a unique function of $\lambda$, which only mildly 
splays out 
if $\phi_c$ is decreased. Both quantities furthermore arrest at 
finite values
in the limit of $\xi_0\to\infty$, since the relevant length scale is
then the colloid size.

\begin{figure}[h]
\centerline{{\epsfysize=9.cm 
  \epsffile{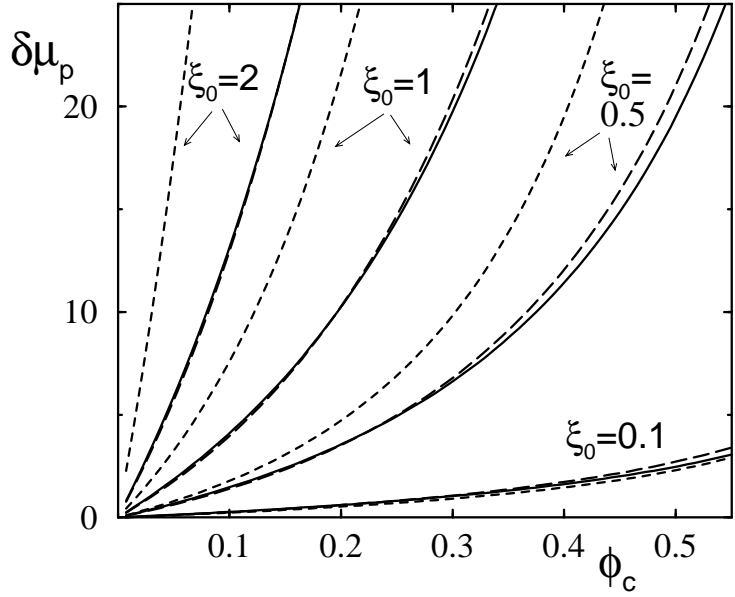}}}
\caption{Polymer excess chemical potentials at infinite dilution,
$\delta\mu_p|_{\varrho_p=0}$ (in units of $1/(N\beta)$)
as functions of the colloid packing fraction
for four size ratios $\xi_0$ as labelled. The solid lines give the
results from long wavelengths,
 Eq. (\protect\ref{m8}), the dashed lines the
corresponding ones from local packing, Eq. (\protect\ref{m10}). The 
short dashed line compares the equivalent PY result for inserting a sphere
of radius equal to $R_g$ as used in the phantom sphere approach of
Ref. \protect\cite{Lekkerkerker92}.
\label{fig9}}
\end{figure}

If the solvation free energy for adding non-adsorbing
 polymers to a fluid of spheres is considered,
\gls{m8}{m10}, then the ideal gas limit result discussed in section IV only
holds for point polymers, $\xi_0\to0$. As expected and shown in Fig. 
\ref{fig9}, 
it becomes more difficult to add larger polymers because less free volume
is available for them. The compressibility expression, 
\gl{m8} can be integrated analytically yielding:
\beqa{r3} &
N \beta \delta \mu^{\rm (c)}_p|_{\varrho_p=0}  =
- \ln{(1-\phi_c)} +
\frac{6\phi_c\xi_0(1+\frac{4}{\lambda_1})}{\sigma_c(1-\phi_c)}  
&\nonumber\\& +
\frac{6\phi_c\xi_0^2(2+\phi_c)(1+
\frac{2}{\lambda_1})}{\sigma_c^2(1-\phi_c)^2}
+\frac{6\phi_c\xi_0(\frac{4}{\lambda_1^2}-\frac{2}{\lambda_1})/\sigma_c
}{(1+\lambda_1\frac{\xi_0}{\sigma_c})
(1+\lambda_1\frac{\xi_0}{\sigma_c}-
\phi_c(1-2\lambda_1\frac{\xi_0}{\sigma_c}))} \; ,
& \eeqa
 which was compared to the virial route in Fig. \ref{fig3}:
For larger polymer coils, the added macromolecule looses 
conformational entropy
when squeezing into the fluid interstitials and thus the free energy cost
increases. Yet, its increase is not as rapid as if the polymer was a sphere
because the polymer chain can rearrange. This becomes especially important
for large polymers because they can wrap around the fluid particles.
 Therefore, if \gl{r3} is compared
with the corresponding result from scaled particle or PY-theory for the 
addition of spheres with size $R_g$ to a sphere fluid \cite{Lekkerkerker92}
then  two qualitative differences appear for large $R_g\gg\sigma_c$. First,  
 only a quadratic scaling with the polymer size results, 
\beq{r3.5}
\beta \delta \mu^{\rm (c)}_p|_{\varrho_p=0}  \to 
\frac{6\phi_c(2+\phi_c)(1+\frac{2}{\lambda_1})}{(1-\phi_c)^2}
\frac{\xi_0^2}{N\sigma_c^2} \fur
\xi_0\to\infty \; ,
\eeq
whereas for (large) added spheres the chemical potential scales with the volume
of the added species.
The behaviour in \gl{r3.5} is connected to the large-$\xi$ scaling
of the chemical potential for adding spheres to a polymer solution, discussed
in \gls{e5}{e6}, because both quantities are determined from the
polymers' ability to deform around a particle. In \gl{r3.5} this explains
why the rhs becomes independent of the degree of polymerization and 
linearly dependent on the colloid size.
Second, the increase with particle fluid density is weaker than the 
corresponding sphere mixture result, which would predict a cubic divergence
for $\phi_c\to1$. Thus the difference of both approaches becomes more important
at higher sphere concentrations.
Note that the free volume expression used Ref. \cite{Lekkerkerker92} 
is connected to $\delta \mu_p|_{\rho_p=0}$ via
$\alpha = V_f/V = e^{-N\beta \delta \mu_p|_{\rho_p=0}}$. The input to this 
approach, the chemical potential for adding a sphere of radius $R_g$ to
a hard sphere solution, is included in Fig. \ref{fig9}, and for large polymers
strongly overestimates the free energy cost for insertion.

\begin{figure}[h]
\centerline{{\epsfysize=9.cm 
  \epsffile{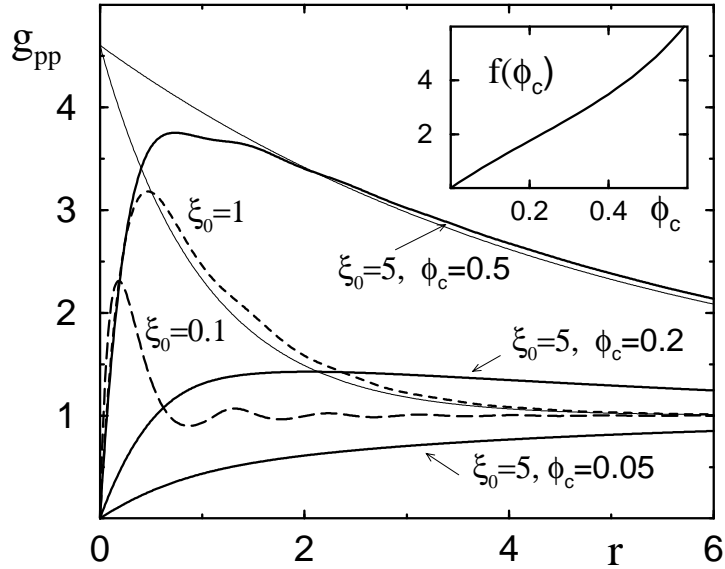}}}
\caption{Polymer-polymer  pair correlation function, $g_{pp}(r)$, 
for dilute polymer molecules immersed in a hard sphere solution; 
the colloid packing fractions $\phi_c$ and size ratios $\xi_0$ are
the same as in Fig. \protect\ref{fig8} and as labelled ($\phi_c=0.5$ for
$\xi_0=$ 0.1 and 1). The thin lines present the asymptote for large polymers,
Eq.  (\protect\ref{r4}), evaluated for $\xi_0=5$ and 1 at $\phi_c=0.5$,
 while the inset shows the intermolecular
polymer segment contact value $f(\phi_c)$, which determines this asymptote.
\label{fig10}}
\end{figure}

The pressure exerted by the particle fluid on the polymer also manifests
itself in the dilute limit intermolecular packing of the segments of 
two polymers which is described by $g_{pp}(r)$ for $\varphi_p\to0$; 
examples are shown in Fig. \ref{fig10}.
The potential-of-mean force between segments is given by $- k_BT
 \ln{g_{pp}(r)}$, while its polymer molecule analog is of order $N^2$ times
 larger.
The slight repulsion of polymers in solution which causes the correlation 
hole is overcome at the distance connected to the depletion layer in
$g_{cp}$; as seen by comparison with  Fig \ref{fig8}.
For larger distances the polymer segments
are pushed together and therefore pack more densely than random. 
This corresponds to a strong tendency of
 polymers to cluster and interpenetrate even though $\varrho_p \ll
 \varrho_p*$.   For
small  polymers an oscillatory pattern develops connected to the distance
of voids between the particles. 
 For larger polymers, this segmental layering
is smeared out and after an initial rise $g_{pp}(r)$ monotonically decays
to its random value. In the limit where the particles are much smaller 
than 
the  two polymeric molecules, but of course still much larger than the 
segmental repeat unit size, the two polymers entwine strongly.
For $\xi_0\gg 1$ and $\varphi_p\to0$:
\beqa{r4} &
g_{pp}(r) \to 1 + ( f(\phi_c) - 1 ) \; e^{-r/\xi_0} & \nonumber \\ &\wer
f(\phi_c) = \frac{\phi_c(6 \lambda_1 + 1 - 
4 \phi_c)}{2(1-\phi_c)(1+2\phi_c)}
\; ,&
\eeqa
follows from \gls{dp2}{dp8}. Within the coil radius, for $r\le\xi_0$,
the distribution of  segments from two chains, is almost constant and 
does not exhibit the self-similar power-law behaviour, $1/r$, 
of open (Gaussian)  fractals. 
Only for distances larger than the coil  size, $r>\xi_0$,
does the intermolecular segment distribution decay to uncorrelated  packing. 
The probability of segments of different polymer chains to be close,
as measured by the pair correlation function contact value 
$f_\infty(\phi_c)$,
is determined by the particle packing fraction, and for $\phi_c\ge0.11$ 
becomes (much) higher than the bulk density. This increase is a precursor
of the demixing transition at finite polymer concentrations \cite{Fuchs00}.
For $\phi_c \approx 0.5$,
 the density of segments from other chains within a $R_g$ distance of
 a tagged segment is more than an order of magnitude greater than in the
 absence of colloids. This may have significant consequences for
 intermolecular processes such a chemical reactions or energy transfer
 between polymers added  at dilute levels to colloidal suspensions or porous
 materials.

\begin{figure}[h]
\centerline{{\epsfysize=9.cm 
  \epsffile{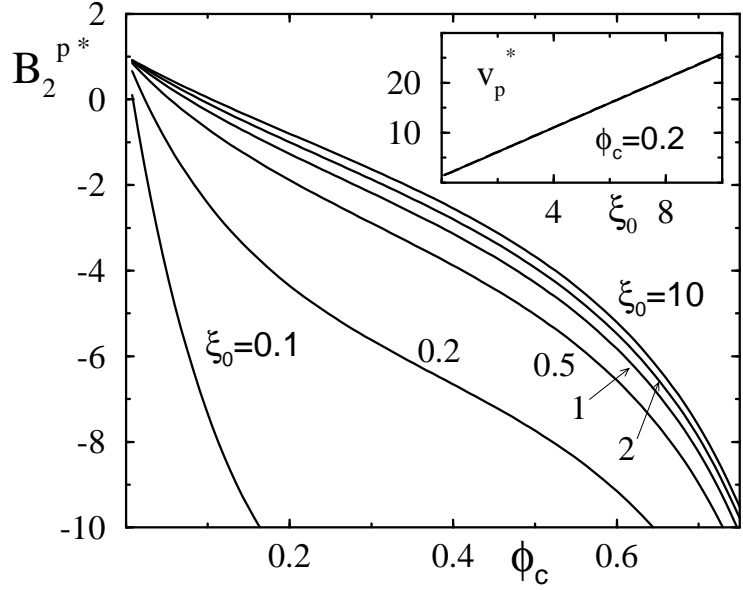}}}
\caption{Normalized polymer second virial coefficient $B_2^{P*}=
B_2^p/(4\pi\xi_0^3)$ as a function of the
colloid packing fraction for various size ratios $\xi_0$ as labelled.
The inset shows the normalized
polymer intermolecular  excluded volume parameter,
$v_p^* = v_p^{\rm excl.} \xi_0/(8\pi l_p^4)$, as a function
of the polymer size, $\xi_0$, and the (indistinguishable) approximation 
Eq.  (\protect\ref{r5}); for the inset,
 the colloid packing fraction is $\phi_c=0.2$.
\label{fig11}}
\end{figure}

The intermolecular excluded volume parameter, defined as
$v^{\rm excl.}_p(\phi_c)= -
\hat{c}_{pp}(0)$, or the related polymer molecule second 
virial coefficient, $B_2^p$, determine
the importance of the excluded volume interaction on the polymer packing on
local and 
macromolecular distances, respectively. Results are shown in Fig. \ref{fig11}.
On local distances, it appears plausible that
 a polymer chain should experience the identical
steric repulsion from a segment of its own backbone  as from another of the
chemically identical macromolecules. Thus, the result in \gl{ap11}
can be used to discuss the effective excluded volume parameter
 induced by the depletant
particles. For vanishing colloid concentration, where $v^{\rm excl.}_p$ and
$B_2^p$ are directly related,
 the interpretation of
polymer chains as repulsive spheres is recovered in PRISM because 
$v^{\rm excl.}_p(\phi_c)\propto 1/R_g$ and therefore $B_2^p \propto R_g^3$.
Note, that our present model
 thus does not describe $\Theta$-solvents, where
$B_2^p=0$, even though a Gaussian intramolecular structure was 
assumed a priori,
but rather describes polymers in athermal solutions with the technical
simplification
 to treat the polymer chains as random walks. For 
results with self-avoiding statistics, see, e. g. Refs. 
\cite{fractal,FuchsMueller}. The result in \gl{ap11}, which to 
a very good
approximation simplifies to
\beq{r5}
v^{\rm excl.}_p(\phi_c)= \frac{8\pi l_p^4}{\xi_0} \;
\left( \frac{1}{1-\phi_c} + \frac{3(2+\lambda_1)^2}{\lambda_1^2}\,
\frac{\phi_c\xi_0/\sigma_c}{(1-\phi_c)^2} \right) \; ,
\eeq
indicates that the effective
 local segmental interaction stays repulsive  for all  size 
ratios and particle fluid concentrations. Asymptotically for small
depletant particles, it becomes  independent of the macromolecular size, 
and saturates to a positive value which increases with $\phi_c$.
 The depletion attraction induced by the
polymer on the  particles overcomes the effectiveness of 
the particles to push polymer segments together on a local scale. 
This is found even though,
mesoscopically, two polymers are induced to intertwine, and for somewhat
higher polymer concentration, phase separation into a polymer gas and 
polymer
fluid phase sets in \cite{Fuchs00}.
On macromolecular distances, however, the effective polymer-polymer molecule 
interaction can become very attractive if the particle density is 
high, or the colloidal particles are large:
\beq{r6}
B_2^p \to \left\{\begin{array}{ll}
-\frac{\pi\sigma_c^3}{12} \left[\frac{(1-\phi_c)^2\phi_c}{(1+2\phi_c)^2} +
\frac{24 \phi_c ( 1- \phi_c)}{(1+2\phi_c)^2} \frac{\xi_0}{\sigma_c}
 \right] & \mbox{for }\;
\xi_0 \to 0 \; ,\\
& \\
\frac{4\pi\xi_0^3}{(1+2\phi_c)} \left[ 1 - \frac{36\phi_c}{(1-\phi_c)}\,
\frac{1+\lambda_1+\frac 13 \lambda_1^2}{\lambda_1^3} \right] & 
\mbox{for }\;
\xi_0\to\infty\; .
\end{array}\right. 
\eeq
For small polymers, the result known from PY theory for hard
sphere mixtures is recovvered. It is negative and
 describes the tendency
of the polymer point particles to cluster. Increasing the polymer size 
somewhat increases this tendency.
For large polymers, however, a finite colloid density is required in 
order for
the colloid induced attraction to overwhelm the 
segmental repulsion. The virial coeficient becomes negative only for 
 $\phi\ge 0.108\ldots$ for $\xi_0\gg1$.

\section{Conclusions}

We  have studied athermal
colloid and polymer mixtures and considered the dilute limit of
one species using a binary mixture approach which treats the  hard
spheres and Gaussian polymer coils on an equal footing. The
macromolecular liquid state theory uniquely addresses the structural
correlations over the wide range of length scales from
the polymer repeat unit size to the molecular sizes.
Packing of hard spheres is handled using the reliable
Percus-Yevick approximation. Polymer specific effects are captured
which arise from the ability of the polymer coils to deform close to
and around the particles. Also, the appropriate polymer correlation
length appears in the description because the formation of a polymer
mesh is captured.
The polymer specific effects become important as
soon as the size of the coils is not negligible relative to the
colloid radius.

The comparison  of the m-PY approximation with rigorous field
theoretic results in the dilute
colloid limit presented in section V serves to validate our approach which
offers the unique possibility of extension to higher densities. 
Interesting and novel predictions arise for the colloid induced pair
interaction of dilute polymers which are much larger than the
particles. Strong interpenetration is predicted (see Fig. \ref{fig10})
finally leading to fluid-fluid phase separation 
\cite{Fuchs00}.

In this context it is important to stress that the approximation of
Gaussian single chain correlations is done for purely technical
reasons in order to achieve analytical results. Of course this entails
that the majority of the scaling predictions derived in the present
and previous work \cite{Fuchs00} bears the wrong exponents if
applied to polymers in a good solvent.  Intramolecular excluded volume,
thermal attractions and a self-consistent determination of $\omega(q)$
 will be included in future numerical studies, and in a number of
 cases the corrected scaling predictions have already been pointed out 
\cite{fractal,FuchsMueller,Fuchs00}.

Colloidal dispersions containing added free polymer  are often
described by integrating out the polymer degrees of freedom in order
to derive an effective colloidal Hamiltonian
 \cite{Dijkstra99}.  However, for  interacting polymers
 there exists no small parameter, and the induced
many-particle interactions among the colloids do not generally
 terminate at a
pair-wise description, and the polymer density dependence of the
parameters in the effective interactions can not be neglected. Both
assumptions enter the description of polymer-colloid mixtures using
effective pair-potentials like the Asakura-Osawa model
\cite{Asakura54,Gast83}, and are justified only for low concentrations of
polymers much smaller than the colloid particles, and preferrably
dissolved in $\Theta$-solvents which mimimize the excluded volume
interaction. The discussion of semi-dilute polymer solutions in
section V  clearly identifies the importance of the density
dependent polymer blob size or mesh-width, which also appears in
$\Theta$-solvents \cite{degennes}. The importance of induced many-body
interactions, which naturally are contained in the presented binary
mixture approach, was discussed in
Refs. \cite{Fuchs00,frenkel}. Treatments of binary mixtures
of spheres have been used to describe the colloid-polymer mixtures,
 replacing the polymer coils by spheres of equal or
similar size and neglecting their direct interactions (``phantom
sphere models'') \cite{Lekkerkerker92,Louis99}. These capture induced
many-body interactions but neglect the deformability of polymers
around particles and the crossover of the relevant polymer length
scale from the coil size to the blob size when reaching semi-dilute
concentrations. Whereas these effects had been well appreciated when
adding dilute particles to polymer solutions, their study in
concentrated particle solutions in the present approach clearly
reveals their importance whenever the size of the polymer coils is not
negligible. Their description in recent ``soft-colloid'' approches to
polymers is yet unclear \cite{Louis,Eurich01}.  An
interesting quantity in this context is the free energy cost
 for adding polymers to hard
sphere fluids, because this is the central input for the widely used
free volume theory of Lekkerkerker et al. \cite{Lekkerkerker92}. As
Fig. \ref{fig9} shows polymer deformability leads to strong deviations
compared to results based on replacing polymer coils by hard spheres.

A central ingredient of the present approach consists in enforcing
thermodynamic consistency for the polymer insertion chemical potential
$\delta\mu_p|_{\rho_p=0}$. The length scale $\lambda$ over which
polymer segments are allowed to rearrange close to the particles is
determined from equating expressions for $\delta\mu_p$ from long
wavelength and from local packing information. Although the latter
compares favourably with exact results for dilute systems,
quantitatively we consider the former coarse-grained approach
(``compressibility route'') to be more reliable, and also
preferentially use it for other thermodynamic quantities within the
present approach. Our reasoning rests on three observations: $(i)$ The
used scaling law approach (thread model)
to semi-dilute polymer solutions is
coarse-grained a priori so a zero wave vector thermodynamic route is
natural. Recently it has been connected to a
self-consistent Gaussian field theory \cite{Chandler93}. $(ii)$ The
compressibility route leads to thermodynamic results far less
dependent on the specific closure as can be seen by comparing the new
m-PY with previous PY results
\cite{avik,avik3,Kulkarni99}. $(iii)$ In a microscopic
calculation the universality, in field-theoretic sense, of the
long-wavelength structure could be shown explicitly
\cite{threada}, in contrast to the local structure route where
  microscopic parameters
remained as prefactors. In the present context this implies that a
full PRISM calculation could lead to a local matching length in
\gl{l2} dependent on the excluded volume size.  Thermodynamic
consistency, nevertheless, should prove a useful concept also for the
description of other systems, like polyelectrolyte mixtures
\cite{Ferreira00}, where PRISM-based numerical  approaches have already given
interesting results.

\acknowledgments{ We gratefully acknowledge fruitful discussions with
L. Belloni, S. Egelhaaf, E. Eisenriegler, M. Schmidt and R. Sear. 
 M.~F. was supported by the Deutsche
Forschungsgemeinschaft under Grant No. Fu 309/3 and through the SFB 563.
K.~S.~S. was supported by the U.S. DOE Division of Materials Science Grant
No. DEFG02-96ER45539 through the UIUC Materials Research Laboratory.}

\appendix

\section{Factorization of the integral equations}

Three-dimensional Fourier transformations leading to $\hat{f}(q)$ of functions
$f(r)$ depending on the radius $r$ only,
 shall be simplified to one-dimensional ones by:
\beq{a1}
\hat{f}(q) = 2 \pi \int_{-\infty}^\infty dr\; e^{i q r } \; \tilde{f}(r) \wer
\tilde{f}(r) = \int_{|r|}^\infty ds \; s\, f(s) \; .
\eeq

\subsection{The limit $\phi_c\to 0$}

Inserting \gl{dc1},  the left equation  of \gl{dc3} becomes:
\beq{ac1}
\hat{h}_{12} = \xi^2 \frac{\hat{\bar c}^s_{12}}{(1+\lam^2q^2)(1+\xi^2q^2)} \; ,
\eeq
which suggests an ansatz for $\bar{c}^s$ of the form:
\beq{ac2}
\hat{\bar c}^s_{12}(q) = 2 \pi e^{-iq/2} ( u_a + i \lam q v_a ) +
(1+ i q \lam ) ( 1 + i q \xi ) \hat{q}_{12}(q) \; .
\eeq
This shifts the problem of finding
 $\bar{c}^s_{12}(r)$ for $-\frac 12 \le r \le 
\frac 12$, to the problem of finding the (Wiener-Hopf factor) function 
$q_{12}(r)$, where
\beq{ac3}
\hat{q}_{12}(q) = 2 \pi \int_{-\frac 12}^{\frac 12} dr \;
e^{i q r} q_{12}(r) \; ,
\eeq
and $q_{12}(r)=0$ elsewhere.
 From the required symmetry,
$\hat{\bar c}^s_{12}(q) =  \hat{\bar c}^s_{12}(-q)$, and  \gl{ac2}, 
follows the continuity of $q_{12}$ at the upper boundary:
\beq{ac4}
q_{12}(\frac 12) = 0\; .
\eeq
Inserting \gl{ac2} into \gl{ac1} leads to
\beq{ac5}
(1- i q \lam ) ( 1 - i q \xi ) \hat{h}_{12}(q) - \xi^2 \hat{q}_{12}(q) =
 \frac{2 \pi \xi^2 e^{-iq/2} ( u_a + i \lam q v_a ) }{
(1+ i q \lam ) ( 1 + i q \xi )}\; ,
\eeq
where by simple inspection the right hand side has no pole in the lower 
complex $q$-plane, $\Im q < 0$. Thus for $r> - \frac 12$ the Fourier-back 
transform of \gl{ac5}, remembering \gl{a1}, can be performed and leads to
\beq{ac6}
( 1 + \lam \partial_r )( 1 + \xi \partial_r ) \tilde{h}_{12}(r) =
\xi^2 q_{12}(r) \fur r > - \frac 12 \; .
\eeq
For $r \ge \frac 12$, this leads to \gl{dc4} and in the overlap region, the 
excluded volume condition gives:
\beq{ac7}
r + \lam + \xi = \xi^2 q'_{12}(r) \fur - \frac 12 < r < \frac 12 \; ,
\eeq
which can easily be integrated using \gl{ac4} and provides the initial
condition for the derivatives of $f_{12}$ at $r=\frac 12$ mentioned in the 
main text. Also the zero wave vector value of the direct correlation function
in \gl{dc5}  follows from \gl{ac2} and the symmetry requirement on
$\hat{\bar c}^s$, which fix:
\beq{ac8}
u_a = -\frac{\xi^2+2\lam\xi(1+\xi)+\lam^2(1+2\xi)}{\xi^2}\quad , \quad
v_a = - \frac{\xi+\lam}{\xi}\; .
\eeq
This  also predicts a Dirac-delta
spike repulsion situated at the surface of the excluded region,
which results in the large-$q$ behavior:
\beq{ac10}
\hat{\bar c}^s_{12}(q\to\infty) \to - 2\pi \frac \lam \xi ( 1 + 2 \lam
+ 2 \xi)\; \cos{\frac q2} \;.
\eeq

Inserting the ansatz \gl{ac2} at $-q$ into the second equation of \gl{dc3}
leads to:
\beq{ac11}
\hat{h}_{22}-\hat{c}_{22} = \frac{\hat{h}_{12}(q)\varphi_p}{2\pi\xi_0} (
\frac{2 \pi e^{iq/2} ( u_a - i \lam q v_a )}{1+\lam^2q^2} +
\frac{( 1 - i q \xi ) \hat{q}_{12}(-q)}{
(1 + i q \lam ) } )\; ,
\eeq
where \gl{ac5} shows that the last term on the rhs
has no pole for $\Im q < 0 $ 
except for a single pole at $q=-\frac i\lam$. Introducing the constants
\beq{ac12}
z = \frac{\lam(u_a+v_a)}{2(\lam+\xi)} \quad ,\quad
u = z - u_a \quad \mbox{and}\quad v = - \frac \xi \lam \; z \; ,
\eeq
also the first term on the rhs of \gl{ac11}
can be decomposed into poles and zeros
 in different
half-planes because of:
\beq{ac13}
\frac{( u_a - i \lam q v_a )}{1+\lam^2q^2} =
z \frac{(1 - i \xi q)}{1+i\lam q} -
\frac{u - i \lam q v}{1-i q \lam} \;
\eeq
On the lhs of \gl{ac11}, 
as $c_{22}(r)$ vanishes for $r>1$, its  factorization can be achieved
using a constant $w$ and an undetermined 
function $q_{22}(r)$ which vanishes outside the range $0\le r \le 1$, and 
whose Fourier transform is given by:
\beq{ac14}
\hat{q}_{22}(q) = 2 \pi \int_0^1 dr\; e^{iqr} q_{22}(r) \; .
\eeq
Respecting that $\hat{c}_{22}$ is a symmetric function of $q$, 
the following ansatz is required:
\beqa{ac15} &
\hat{c}_{22}(q) = \frac{2\pi w}{1+\lam^2q^2} + 
\frac{\hat{q}_{22}(q)}{1-i\lam q} +
\frac{\hat{q}_{22}(-q)}{1+i\lam q} & \nonumber \\ & -
\frac{\varphi_p\xi^2}{2\pi\xi_0} \frac{
(\hat{q}_{12}(q)+ 2\pi z e^{-iq/2})(\hat{q}_{12}(-q)+ 2\pi z e^{iq/2})
}{1+\lam^2q^2} \; ,&
\eeqa
where the requirement of  absence of a term of the form $1/q^2$ for 
$q\to\infty$, which would correspond to a divergence $c_{22}(r\to0)\sim 1/r$,
 determines the constant $w$ from the value of $q_{22}$ at $r=0$:
\beq{ac16}
q_{22}(0) = \frac{w}{2\lam} - \frac{\varphi_p \xi^2}{2\lam\xi_0} \; z^2 \;.
\eeq

The ansatz \gl{ac15} is useful because it cancels a number 
of the poles in the lower complex $q$-half plane  in \gl{ac11}.
The required short range of the direct correlation function, $c_{22}(r>1)=0$,
however, at first sight is violated by \gl{ac15}, which appears to indicate
an exponential tail, $r c_{22}(r) = {\rm const.}
  e^{-|r|/\lam}$ for $|r|>\frac 12$. 
 Requiring the constant to be zero, i.e. requiring the residues
 of the rhs of \gl{ac15} at $q=\pm i/\lam$ to vanish,
fixes the initial value of the factor function:
\beqa{ac17} 
q_{22}(0) & = & \frac{-1}{\lam} \int_0^1 dr\; e^{r/\lam} q_{22}(r) +
\frac{\varphi_p \xi^2}{8\pi^2\lam\xi_0} (
\hat{q}_{12}(\frac{-i}{\lam}) \hat{q}_{12}(\frac{i}{\lam}) \nonumber\\ 
& + & 2 \pi z e^{1/(2\lam)} \hat{q}_{12}(\frac{-i}{\lam}) 
+ 2 \pi z e^{-1/(2\lam)} \hat{q}_{12}(\frac{i}{\lam}) ) \; , 
\eeqa
which is a linear equation in $q_{22}(0)$. Unfortunately, the full expression
for $q_{22}(0)$ turns out rather complicated and only simplifies if 
the fact that $\lam\ll 1$ holds is recalled in order
to neglect corrections of the order ${\cal O}(e^{-1/\lam})$.

The factorization of $\hat{c}_{22}$ expressed in \gl{ac15} simplifies the 
Fourier-back transformation of \gl{ac11}, as the poles in the lower
complex $q$-plane were identified in \gls{ac13}{ac15}.
Thus for $r>0$ one finds after Fourier-back transforming  that
\beq{ac18}
(1+\lam \partial_r) \tilde{h}_{22}(r) + \frac{\varphi_p}{\xi_0}
(u + v \lam \partial_r) \tilde{h}_{12}(r-\frac 12) = q_{22}(r) \; ,
\eeq
holds, which proves \gl{dc6} and, together with \gl{ac17},
also proves \gl{dc7}, because of $g_{22}(1) = q_{22}(1)/\lam$.
In the overlap region, it  leads to:
\beq{ac19}
q'_{22}(r) = r + \lam + \frac{\varphi_p}{\xi_0} ( u r - \frac u2 + v \lam )
\fur 0 < r < 1 \;.
\eeq
The changes in the contact value upon adding
more colloidal particles, i.e.
 $(\partial g_{cp}/\partial \varrho_c)|_{\rho_c=0}$, can  be determined in an
explicit form  from the Wiener-Hopf factorization of the full non-linear 
equations  \cite{inpreparation}.

\subsection{The limit $\varphi_p\to 0$}

For $\varphi_p\to0$, the colloid structure factor agrees with the PY-solution
for hard spheres, which can be written in terms of Baxter's factorization
function as:
\beqa{ap1} 
\hat{\bar{S}}_{22}^{-1} &=& (1-\hat{q}_{22}(-q))(1-\hat{q}_{22}(q)) \wer
 \nonumber\\ 
\hat{q}_{22}(q) &=& 2\pi \int_0^1 dr\, q_{22}(r) \; e^{i q r} \nonumber\\ &=&
2\pi \int_0^1 dr\; ( \frac A2 (r^2-1) + B ( r - 1 ) )\; e^{i q r}\; ,
\eeqa
with the coefficients: $A=\varrho_2 \frac{1+2\phi_c}{(1-\phi_c)^2}$ and
$B=\varrho_2\frac{-3\phi_c/2}{(1-\phi_c)^2}$. 
Correspondingly the pair correlation function satisfies:
\beq{dp1h}
f_{22}(r) 
- 2\pi  \int_{0}^{1} ds \; q_{22}(s) f_{22}(r-s)
=  0
 \fur r\ge 1 \; .
\eeq
The first of \gl{dp1}  can be 
simplified with the ansatz:
\beqa{ap2} 
\hat{\bar c}^s_{12}(q) &=& (1+i q \xi_0) ( 1 + i q \lam ) \hat{q}_{12}(q) 
 \nonumber \\  &+&
e^{-i q/2} ( 1 - \hat{q}_{22}(q) ) 2\pi ( u_b + i q \lam v_b )\; , 
\eeqa
as from Baxter's solution it is known that $(1-\hat{q}_{22}(q)) 
(1+\varrho_2 \hat{h}_{22}(q))$ 
has no poles for $\Im q < 0$. 
The function $q_{12}(r)$
again is assumed to vanish outside the overlap region such that \gl{ac3} holds.
The required symmetry of $\hat{c}_{12}(q)$ and the known properties of
$q_{22}$ further show that \gl{ac4} holds and that 
$v_b=\xi_0 q_{12}(-\frac 12)$
and $u_b=-2\pi q_{22}(0)\lam v_b +v_b(1+2\lam+\lam/\xi_0)$.
 Inserting \gl{ap2} into
\gl{dp1} and closing the Fourier-integrals in the lower complex-$q$ half 
plane thus leads for $r>-\frac 12$ to
\beqa{ap3} &
(1+\xi_0 \partial_r) ( 1 + \lam \partial_r) \tilde{h}_{12}(r) & \nonumber \\
& =
\xi_0^2 q_{12}(r) + 12 \phi_c \xi_0^2 \int_{-\frac 12}^{\frac 12} ds\;
q_{12}(s) \tilde{h}_{22}(r-s) \; .&
\eeqa
This leads to \gl{dp2} for $r>\frac 12$ as $f_{ij}(r) = - \partial_r 
\tilde{h}_{ij}(r)$ for $r>0$. Within the overlap region the equation 
determining $q_{12}$ follows:
\beq{ap4}
r + \xi_0 + \lam = \xi_0^2 q_{12}'(r) + 12 \phi_c \xi_0^2
\int_{-\frac 12}^{\frac12} ds \; q_{12}(s) (r-s) \; ,
\eeq
which is solved by the result given in \gl{dp4} with parameters:
\beqa{dp5} 
a &=& \frac{1 - \phi_c\,\left( 1 - 6\,\lam - 
       6\,\xi_0 \right) }{{\left( 1 - \phi_c
        \right) }^2\,{\xi_0}^2}  \\
b &=& \frac{\lam + \xi_0}
    {\left( 1 - \phi_c \right) \,{\xi_0}^2}
      \; .
\eeqa
From the discontinuity
of $q_{12}'$ at $r=\frac 12$, the initial condition \gl{dp3} results.
Solving for the parameters
\beqa{ap5}
u_b & = & 
-\frac{\left( \lam + 
        \xi_0 \right) \,
      \left( \xi_0 - \phi\,\xi_0 + 
        \lam\,
         \left( 1 + 2\,\xi_0 - 
           \phi\,\left( 1 - 4\,\xi_0 \right) 
           \right)  \right) }{{\left( 1 - \phi \right) }^
       2\,{\xi_0}^2}
\; , \nonumber\\
v_b & = & \frac{\lam + \xi_0}
  {-\xi_0 + \phi\,\xi_0} \; ,
\eeqa
one can obtain the \gls{dp6}{dp7} from \gl{ap2}.
The second of \gl{dp1} can be rewritten upon insertion of \gl{ap2}
\beqa{ap6} &
\hat{h}_{11} - \frac{\xi_0^4}{(1+q^2\xi_0^2)^2} \frac{\hat{c}_{11}}{l_p^4} =
\frac{6\phi_c \xi_0^2}{\pi} \; \frac{\hat{h}_{12}(q)}{(1-i q\lam)(1-iq\xi_0)}
& \nonumber \\ & \times
\left(  \hat{q}_{12}(q) + \frac{2\pi e^{-iq/2} 
(1-\hat{q}_{22}(q))(u_b+iq\lam v_b)}{
(1+i q\lam)(1+iq\xi_0)} \right) \; ,
\eeqa
Considering the explicit expression for $\hat{h}_{12}(q)$ from Eqs.
(\ref{dp1},\ref{ap1},\ref{ap2}):
\beqa{ap7} &
\hat{h}_{12}(q) = 
\hat{h}_{12}(-q) =
 \frac{\xi_0^2  \hat{q}_{12}(-q)}{
(1-\hat{q}_{22}(-q))(1-\hat{q}_{22}(q))(1+i q\lam)(1+iq\xi_0)} 
&\nonumber\\ 
 &+ \frac{2\pi \xi_0^2  e^{iq/2} (u_b-iq\lam v_b)}{ (1-\hat{q}_{22}(q))
(1+q^2\lam^2)(1+q^2\xi_0^2)}  \; , &
\eeqa
one recognises that the first term of \gl{ap7} multiplied with
$(1-\hat{q}_{22}(q))$ contributes no  poles
in the lower $q$-half plane. Thus the poles for $\Im q < 0$, which are the
 only ones
contributing  for $r>0$,  from the second part of \gl{ap6} can be identified:
\beqa{ap8} &
 \frac{12 \phi_c \xi_0^2  e^{-iq/2} 
(1-\hat{q}_{22}(q))(u_b+iq\lam v_b)}{
(1+i q\lam)(1+iq\xi_0)} \; \hat{h}_{12}(q) & \nonumber \\ &=
 \frac{2\pi\lam^2 z_\lam}{1-iq\lam} + \frac{2\pi \bar{z}_{\xi_0}\xi_0^2}{
1-iq\xi_0} + \ldots \; ,&
\eeqa
where the parameters are given by
\beqa{ap9}
 z_\lam & = & 
\frac{3\phi_c \xi_0^4 \lam }{(\lam-\xi_0)(\lam+\xi_0)^2} 
( u_b^2-v_b^2 ) \; , \nonumber \\
\bar{z}_{\xi_0} & = & 
\frac{3\phi_c \xi_0^3}{(\lam-\xi_0)(\lam+\xi_0)^2} 
(\lam^2v_b^2-u_b^2\xi_0^2 )\; ,
\eeqa
and $z_{\xi_0} = \bar{z}_{\xi_0}
- \frac{\hat{c}_{11}\xi_0(\lam-\xi_0)}{8\pi l_p^4}$. From \gls{ap6}{ap9} the
\gl{dp8} follows by transformation and differentiation.
The last unknown quantity, $\hat{c}_{11}(q=0)$, can be obtained from \gl{ap6}
because the rhs vanishes as $\cos^2{(q/2)} / q^{8}$, as can be seen from 
\gls{dp1}{ap3}. Therefore, the discontinuity of the third derivative at $r=0$
of the  transform of $\omega_q^2 \hat{c}_{11}$ 
needs to be balanced by $\hat{c}_{11}$:
\beq{ap10}
\partial_r^3 \tilde{h}_{11}(r)|_{r=0} =-
\partial_r^2 f_{11}(r)|_{r=0} =  \frac{\hat{c}_{11}(0)}{4\pi l_p^4}\; ,
\eeq
which together with \gl{dp8} fixes the polymer excluded volume parameter as
given in \gl{ap11}

The change in the colloid-contact value upon adding polymers,
 $(\partial g_{cc}/\partial \varrho_p)|_{\rho_p=0}$, can be
obtained from the equations at finite density and leads to a rather unwieldy
expression \cite{inpreparation}. For this work, only its limits are required:
 $(\partial g_{cc}/\partial \varrho_p)|_{\rho_p=0} ={\cal O}(\xi_0^2)$ for
small $\xi_0$, and neglecting terms ${\cal O}(\lambda^2)$
\beq{ap12}
 \frac{\partial g_{cc}(1)}{\partial \varphi_p}|_{\rho_p=0} \to
\frac{1/(2\xi_0)}{(1-\phi_c)^2} + \frac{2\lambda}{\xi_0} 
\frac{1+2\phi_c}{(1-\phi_c)^3} \fur \xi_0\to\infty \; .
\eeq

\section{Steps in the determination of $\lambda$}
In the limit $\xi_0\gg1$ one can use the 
physically motivated expectation that $\lambda$ becomes small compared to 
$\xi_0$ and is of the order of the colloid diameter. With the
ansatz, $\lambda = \Lambda(\phi_c)$ for $\xi_0\to\infty$, the
expressions in \gls{dp3}{dp6} simplify. With \gl{ap12}, which entails
  a further expansion
in $\Lambda < 1$ up to linear order,  the contact route
expression \gl{m10} becomes:
\beqa{l3} &
N\beta\delta\mu^{\rm (g)}_p|_{\rho_p=0} \to
2 \xi_0^2 \phi_c^{2/3} \int_0^{\phi_c} dx\; 
\frac{1+2x}{\Lambda(x) x^{2/3}(1-x)^2} & \nonumber \\ & +
24 \xi_0^2 \phi_c^{2/3} \int_0^{\phi_c} dx\; x^{1/3} (
\frac{1/2}{(1-x)^2} + 2 (\frac{x}{\phi_c})^{1/3} \, \Lambda(x)  
\frac{1+2x}{(1-x)^3} ) \; .&
\eeqa
An ordinary differential equation for $\Lambda(\phi_c)$
is obtained by considering:
\beqa{l4} &
\frac{\partial}{\partial \phi_c} 
\phi_c^{2/3} 
[\frac{\partial}{\partial \phi_c} ( \phi_c^{5/3} 
\frac{\partial}{\partial \phi_c}  \phi_c^{-2/3}  
\delta\mu^{\rm (g)}_p|_{\rho_p=0} ) & \nonumber \\ &-
\frac{\partial}{\partial \phi_c} ( \phi_c 
\frac{\partial}{\partial \phi_c} 
- \frac 23  ) \delta\mu^{\rm (c)}_p|_{\rho_p=0} ] = 0 \; , &
\eeqa
whose asymptotic solutions are given in \gl{l5}.

\section{Comparison of $B_2^c$ from other closures}

The failure of thermodynamic consistency in the colloid second virial
coefficient $B_2^c$ for situations where the effective colloid pair 
potential is strong, warrants a comparison 
with results obtained with other closures. Because of previous
polymer-colloid mixture work
in Refs. \cite{avik,avik3} we consider two  alternative
approximations: the PY closure for
the colloid-polymer interaction, which corresponds to
 $\lambda=0$, and the HNC closure for
the colloid correlations. Study of the latter is motivated by the
known deficiencies of the PY closure for mixtures of hard spheres, which are
(partially) corrected by the HNC closure \cite{Dickmann97}.

\begin{figure}[h]
\centerline{{\epsfysize=9.cm 
  \epsffile{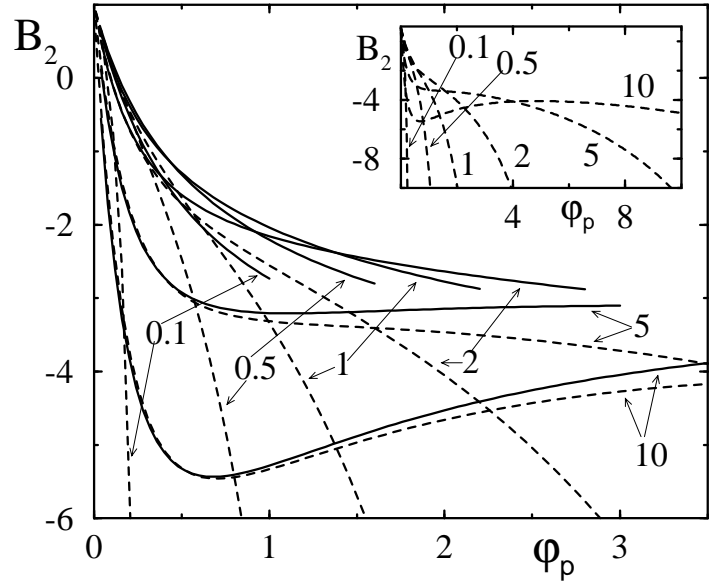}}}
\caption{Colloid second virial coefficient $B_2^{c}$ in units of
$\frac{2\pi}{3}$ as a function of the
polymer concentration for various size ratios $\xi_0$ as labelled.
The solid lines
 are the compressibility results from Fig. \protect\ref{fig7}
whereas the dashed lines give the corresponding  HNC results.
The inset presents the latter in a larger window, showing their
drop to large negative values for polymer concentrations far above the
overlap one.
\label{fig12}}
\end{figure}

The result for the colloid pair correlation function using the
HNC closure can  easily be obtained from the m-PY result via \cite{avik}:
\beq{c1}
g^{\rm HNC}_{cc}(r) = e^{g^{\rm m-PY}_{cc}(r)-1} \fur r>\sigma_c\; ,
\eeq
Because the quantity $\beta W(q)=\varrho_p c^2_{cp}(q) S_{pp}(q)$
in \gl{dc3} can be identified as (Fourier transform of) the polymer
induced potential \cite{Grayce94a},
 the HNC closure agrees with the exact virial expression
for colloidal spheres interacting with this (effective) pair potential
and the $\sigma_c/R_g\to\infty$ blob scaling mean field results
\cite{joanny}.
This holds, because  $\beta W(r)=-g^{\rm m-PY}_{cc}(r)$ 
for $r>\sigma_c$ as follows from \gl{dc3}. Here, $g^{\rm m-PY}_{cc}(r)$ 
is the pair correlation of the m-PY approach presented in the main text.
From Fig. \ref{fig12} one deduces that, as
discussed in section V, the m-PY results are not reliable for
$\sigma_c \gg \xi$ (i.e. small
size ratios $\xi_0$ or higher polymer concentrations) 
because the linearization in \gl{c1} underestimates the
 strong attraction in these cases. For larger polymers,
however, in agreement with the consideration of thermodynamic consistency
in section V,  the PY linearization of \gl{c1} is qualitatively and even
quantitatively appropriate. The minimum in $B_2^c$ at the overlap 
concentration, as well as the semi-dilute scaling law, are present in
the HNC (and $\lambda=0$ PY \cite{avik}
results also. The closer quantitative agreement of the
HNC results for $B_2^c$ with the 
compressibility-route results within m-PY is one of the prime reasons
we favour  this route \gl{dc11},  over the virial type one,  \gl{e10}. 
Moreover, whereas the latter was found to be unphysical in   
the PY closure for the colloid-polymer interaction, $\lambda=0$
\cite{avik,avik3,Kulkarni99}, the former which rests upon
$\hat{c}_{cp}(q=0)$, show, identical qualitative behaviour as the m-PY.
 This can easily be observed by specifying 
the general result \gl{dc5} to $\lambda=0$. The novel m-PY closure
 predicts quantitatively stronger depletion effects than the PY 
and moreover correctly describes the depletion layer.

\end{multicols}

\newpage

\begin{thebibliography}{10}
\bibitem{eisenrieglervergleich}
Failure to recover the $0.40$ exponent is not a failure of PRISM but rather the
  consequence of the use of an Gaussian $\omega(q)$.
\end{thebibliography}
\end{document}